\documentclass[preprint2]{aastex}
\usepackage[]{natbib}
\usepackage{graphicx}
\begin{document}

\title{Description of Stellar Acoustic Modes \\
    Using the Local Wave Concept}

\author{P. A. P. Nghiem}


\affil{Service d'Astrophysique, DAPNIA/DSM/CEA, CE Saclay, F-91191 
Gif-sur-Yvette Cedex}
\email{papnghiem@cea.fr}


\begin{abstract}
An understanding of stellar acoustic oscillations is attempted, using 
the local wave concept in semi-analytical calculations. The local homogeneity 
approximation allows to obtain simplified equations that can nevertheless 
describe the wave behavior down to the central region, as the gravitational 
potential perturbation is not neglected. Acoustic modes are calculated as 
classical standing waves in a cavity, by determining the cavity limits and the 
wave phases at these limits. The internal boundary condition is determined by a 
fitting with an Airy function. The external boundary condition is defined as the 
limit where the spatial variation of the background is important compared to the 
wavelength. This overall procedure is in accordance with the JWKB approximation. 
When comparing the results with numerical calculations, some drawbacks of the 
isothermal atmosphere approximation are revealed. When comparing with seismic 
observations of the Sun, possible improvements at the surface of solar models 
are suggested. The present semi-analytical method can potentially predict 
eigenfrequencies at the precision of $\pm$3 $\mu$Hz in the range 800-5600 
$\mu$Hz, for the degrees $l=0-10$. A numerical calculation using the same type 
of external boundary conditions could reach a global agreement with observations 
better than 1 $\mu$Hz. This approach could contribute to better determine the 
absolute values of eigenfrequencies for asteroseismology.
\end{abstract}


\section{Introduction}
\label{Introduction}
Asteroseismology describes mainly the propagation of matter waves and their 
eigenmodes in stars (or planets) that are considered as cavities. By an inverse 
procedure one can then deduce the physical properties of the layers crossed by 
the waves. It is a very effective tool, and often the unique one to study the 
internal structure of stars.  Since several decades, many intensive works were 
undertaken, especially for the Sun, to calculate precisely the oscillation modes 
and to understand the physical mechanisms involved. Different reviews are 
dedicated to this subject \citep{Cox_1980, Unno_etal_1989, JCD&Berthomieu_1991, 
Turck-Chieze_etal_1993, JCD_cours_1998}. 

Semi-analytical approaches have been developed, essentially based on various 
asymptotic approximations. \citet{Lamb_1909, Lamb_1932}, \citet{Cowling_1941}, 
\citet{Tassoul_1980}, \citet{Leibacher&Stein_1981}, \citet{Duvall_1982}, and 
\citet{Deubner&Gough_1984}, have explored the first-order asymptotic 
approximation, and have helped to interpret the high-order modes measured for 
the Sun. Then with the availability of low-order modes, \citet{Vorontsov_1991}, 
\citet{Lopes&Turck-Chieze_1994},\citet{Roxburgh&Vorontsov_2000}, have taken into 
account the effect of gravity in second-order theories, allowing to better 
describe the propagation in central regions of stars. On another hand, numerical 
calculations are required to reach the high precision in frequency 
determination, needed for inversion of stellar internal layer physics. Several 
high accuracy oscillation codes already exist, offering the computation of 
eigenfrequencies and eigenfunctions \citep{JCD_1982, Unno_etal_1989}. They have 
been widely used to interpret the large amount of solar modes observed from 
Earth or space. 

Just at the beginning of the spatial-asteroseismology era, it is worth assessing 
the progress of these theoretical efforts in the prediction of observations. It 
is interesting to note that the agreement between numerical results and 
observations is remarkable in the range 1000 - 2400 $\mu$Hz where $\nu_{num} - 
\nu_{obs}$ is contained in $\pm$1 $\mu$Hz. At larger frequencies however, the 
disagreement increases gradually, up to 20 $\mu$Hz at 5000 $\mu$Hz. 
Semi-analytical results agree well with numerical ones only in that 
high-frequency range, while their dispersion at lower frequencies stays 
important even when they have been reduced by the second order approximation. 
Moreover, the acoustic cut-off frequency is theoretically predicted around 5000 
$\mu$Hz, while from observations it is rather around 5600 $\mu$Hz 
\citep{Fossat_etal_1992, Garcia_etal_1998}. Those difficulties can be overcome 
in the case of the Sun thanks to the great number of eigenmodes accessible by 
observations. But such difference could be more problematic in asteroseismology 
as only low-degree modes ($l \leq 3$) are available and the stellar radius and 
mass are less known.

The discrepancy between calculation and observation is generally attributed to 
an incorrect description of the outer layers of the current solar models at the 
surface. Several studies have tried to introduce a more relevant model of 
convection via the introduction of a turbulent pressure \citep{Kosovichev_1995}. 
That reduces the discrepancy at high frequencies, but the excellent agreement at 
low frequencies is also deteriorated \citep{Rosenthal_1995, Rosenthal_1999}. The 
role of the chromosphere was also invoked, but no real improvement is yet 
obtained \citep{Wright&Thompson_1992, Vanlommel&Goossens_1999, 
Dzhalilov_etal_2000}.

With the regular progress of observational instruments, these remaining problems 
become more crucial: the theory-observation discrepancies at high frequencies 
and the inaccuracy of semi-analytical methods at low frequencies are up to 100 
times larger than observational errors. So, we suggest to search the reasons not 
only in the solar model at the surface, \emph{but also} in the eigenmode 
calculation itself, especially about the wave description in the solar interior 
and the boundary conditions at the surface. Concepts closer to classical wave 
physics remain to be further explored. It can be considered as a continuation of 
the works of \citet{Leibacher&Stein_1981} and \citet{Duvall_1982}. Some first 
acoustic-frequency calculations was attempted in \citet{Nghiem_2000}, and the 
effect of gravitational potential was taken into account in 
\citet{Nghiem_2003a}. In this paper, we aim to further formalize the concepts, 
by putting a special accent on the internal and external boundary conditions. 
This work is organized as follows. In \S 2, a new p-mode calculation is 
developed. Its approximations and its boundary conditions are proved to be in 
accordance with the JWKB method. In \S 3, the obtained results are first 
compared to present numerical calculations, and some drawbacks of the isothermal 
atmosphere approximation are pointed out. Then in \S 4 comparisons with 
observations suggest a way to improve solar surface descriptions, and at the 
same time to evaluate the potential of the proposed method. Concluding remarks 
are presented in \S 5.

\begin{figure}[htb]
\centering
\plotone {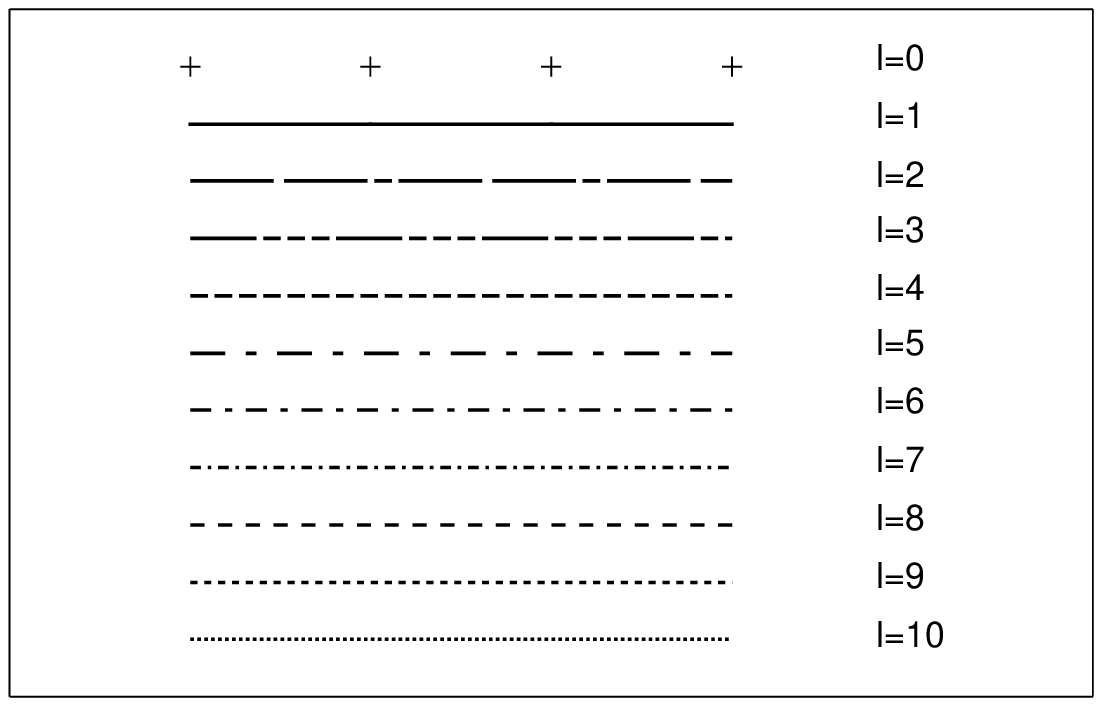}
\caption{Symbols used for the graphic representation of the $l=0$ to 10 modes
in all this paper.}
\label{fig_legende}
\end{figure}

In this paper, the solar model used is the seismic model 
\citep{Turck-Chieze_etal_2001, Couvidat_etal_2003}, computed with the CESAM code 
\citep{Morel_1997}. The numerical results for solar oscillations are obtained 
with the ADIPLS code \citep{JCD_code_1997}. Acoustic mode observations of the 
Sun come from the SoHO spacecraft (MDI data from \citet{Rhodes_etal_1997}; GOLF 
data from \citet{Bertello_etal_2000}, and from \citet{Garcia_etal_2001}.The 
graphic representation for modes of degrees $l = 0$ to 10 follows the symbols 
indicated in Figure \ref{fig_legende}.

\section{The Present Approach}
\label{The Present Approach}
The present study starts from the standpoint that observed solar modes, due to 
their regularity and their precision, are related to waves in its most classical 
sense, excluding shock waves, wave buckets, etc. Basically eigenfrequencies are 
the signature of waves trapped in a cavity. A wave is the propagation of an 
oscillation. An oscillation is a continuous exchange between kinetic and 
potential energy around an \emph{equilibrium state}. Here the notion of 
equilibrium state is the central one. Where this latter ends, there is no 
possible oscillation, thus no more propagation, the wave becomes evanescent. At 
this point, the wave is reflected. That determines the turning point, i.e. the 
cavity limit.

Inside a huge gazeous sphere maintained by gravity like a star, two kinds of 
equilibrium can be considered: the equilibrium between stratums of different 
pressure, and the pressure equilibrium inside each stratum, that is the local 
homogeneity. Every pressure perturbation generates two oscillations: a gravity 
one around an equilibrium stratum, and an acoustic one around a point of local 
equilibrium. The two induced waves do not interact between them, unless their 
frequencies are close to each other. For the Sun, excepted around 500 $\mu$Hz 
where coupling has to be considered, acoustic and gravity waves can be studied 
separately. The present approach  is also applicable for all stars where 
acoustic and gravity frequencies are well separated.

Let us now concentrate only on pure acoustic waves. By definition, such a  wave 
can only propagate in a locally homogeneous environment, or more precisely 
homogeneous as regard to its wavelength. If the pressure scale height can be 
seen as representative of the homogeneity, the propagation condition can be 
expressed as
\begin{equation} \label{prop_cond}
\lambda=\frac{2\pi}{k}< aH_{p(\mathbf{k})}
\end{equation}
where $\lambda$ is the wavelength, $k$ the wavenumber, $H_{p(\mathbf{k})}$ the 
pressure scale height projected on the propagation direction $\mathbf{k}$, and 
$a$ is a constant to be determined. To be exhaustive, this coefficient could be 
considered as the first one of a Taylor expansion in $\lambda$, $H_p$, $H_\rho$, 
or similar parameters. These coefficients should be determined once, by 
experiments on Earth, or spatial experiments, or theoretical studies, but they 
are not specific to a star. When a condition of the type of equation 
(\ref{prop_cond}) is no longer satisfied, the wave can no longer propagate and 
is reflected.

The reflection conditions are different, following the current pressure 
variation. If that happens where the pressure is increasing, for the wave, 
everything is like it encounters a wall, the reflection gets the properties of 
the `pressure wall' type. On the opposite case, that are the properties of the 
`pressure vacuum' type. For this latter, in the absence of any other force, the 
force balance at the boundary requires that the perturbed pressure at the 
discontinuity is
\begin{equation} \label{vacuum_cond}
p'=0.
\end{equation}
One must have a pressure node, corresponding to a displacement antinode, 
because, as seen below, there is a phase lag of $\pi$/2 between them. We will 
see that it is the case at the upper atmosphere limit: the last layer freely 
vibrates under the action of acoustic waves. The situation is exactly the same 
at the free end of a spring laying horizontally on a frictionless surface on 
Earth. Now, in the presence of the gravity force, the force balance requires 
that
\begin{equation} \label{vacuum_cond_under_gravity}
p'+ \rho_0 \int_{}^{}\!\! g'\! dr=0
\end{equation}
where $g'$ is the perturbed gravity, and $\rho_0$ is the equilibrium density 
that can be brought out the integral thanks to the local homogeneity 
approximation. Note that the sign used in this context is opposite to that of 
the continuity equation. As seen later, this will be the situation for the 
radial mode at the internal boundary. This looks like vibrations at the free end 
of a spring vertically hung in the Earth gravity field.

From this viewpoint, the reflexion occurs at a given radius, and the continuity 
condition must be regarded at the two sides of this fixed limit, thus the 
perturbed pressure considered should be eulerian. This is different from the 
perspective where the limit is not fixed in space but follows the displacement 
generated by the perturbation, where the lagrangian perturbed pressure is 
considered.
 
The wave propagation and reflection are so far completely defined in their 
principles. We are in the approximation context of local homogeneity, or pure 
acoustic wave, or of local wave, i.e. the wave is only defined locally. It is 
important to stress that in this approximation, the environment is \emph{not} 
homogeneous, but slowly variable. This choice is only guided by the aim of 
describing what is called a `local-acoustic' wave. The precision reached with 
this approximation is not known a priori, so it will be estimated on the view of 
its expressions and its numerical results, compared to other methods.

\subsection{Wave Propagation}
\label{Wave Propagation}
After a perturbation, the wave propagation is fully described by the four 
eulerian functions $\xi$, $p'$, $\rho'$, $\Phi'$, which are respectively the 
displacement, along with the perturbations of pressure, density and 
gravitational potential. In the hypotheses of adiabacity, sphericity and 
linearity, and when the time, angular, and radial parts have been separated, 
these unknown functions are governed by the following differential equations 
where only remains the radial part (see e.g. \citet{JCD_cours_1998}, p.54):
\begin{eqnarray}
\label{init_eq_1}
\frac{d\xi_r}{dr}&=& \left(-\frac{2}{r}+\frac{1}{\Gamma_1 H_p}\right)\xi_r
    + \left(\frac{l(l+1)}{r^2}-\frac{\omega^2}{c_0 ^2}\right)\frac{p'}{\rho_0 
\omega^2} \nonumber\\
 &&- \frac{l(l+1)}{\omega^2 r^2} \Phi'\\
\label{init_eq_2}
\frac{dp'}{dr}&=& \rho_0(\omega^2-N^2)\xi_r - \frac{1}{\Gamma_1 H_p}p' 
                  + \rho_0\frac{d\Phi'}{dr} \\
\label{init_eq_3}
\rho'&=&\frac{p'}{c_0^2}+\frac{\rho_0}{g_0}N^2\xi_r \\
\label{init_eq_4}
\frac{d^2\Phi'}{dr^2}&=&-\frac{2}{r} \frac{d\Phi'}{dr}
-4\pi G\rho'+\frac{l(l+1)}{r^2}\Phi'
\end{eqnarray}
with r the distance to the stellar center, and $l$ the mode degree. The other 
symbols stand for medium equilibrium quantities: $p_0$, $\rho_0$, $c_0$, $g_0$ 
are the pressure, the density, the sound speed, and the gravity; $H_p$ is the 
pressure scale height, $N$ the buoyancy frequency, and $\Gamma_1$ the adiabatic 
exponent. That is in fact a system of 3 differential equations of fourth order. 
Equation (\ref{init_eq_3}) in $\rho'$ is added only for the sake of 
exhaustiveness.

That system of differential equations can only be solved numerically. An 
analytical solution needs further approximations. Due to that, results obtained 
by numerical codes are more precise and realistic than those of analytical 
methods. Nevertheless, the final result crucially depends on boundary conditions 
that are \emph{not} contained in those equations. The choice of boundary 
conditions comes from physical considerations, and analytical calculations can 
be helpful for that, especially when their results are precise enough to be not 
too far from numerical ones, or in other words, and that is an evidence, when 
the approximations used are not too crude.

In order to find an analytical solution, we propose to use the above suggested 
approximation of local homogeneity (but \emph{not} of homogeneity), where the 
derivatives of the equilibrium quantities are only negligible when compared to 
equilibrium quantities themselves. Only the relative derivatives, namely 
$1/H_p$, $1/H_\rho$, and thus $N^2$, can be neglected. This also mathematically 
justify a validity condition of the type of equation (\ref{prop_cond}). In that 
context, equations (\ref{init_eq_1}-\ref{init_eq_4}) become:
\begin{eqnarray}
\label{acous_eq_1}
\frac{d\xi_r}{dr}&=& -\frac{2}{r} \xi_r
    + \left[\frac{l(l+1)}{r^2}-\frac{\omega^2}{c_0 ^2}\right]\frac{p'}{\rho_0 
\omega^2} \nonumber\\
 &&- \frac{l(l+1)}{\omega^2 r^2} \Phi'\\
\label{acous_eq_2}
\frac{dp'}{dr}&=& \rho_0 \omega^2 \xi_r + \rho_0\frac{d\Phi'}{dr} \\
\label{acous_eq_3}
\rho'&=&\frac{p'}{c_0^2} \\
\label{acous_eq_4}
\frac{d^2\Phi'}{dr^2}&=&-\frac{2}{r} \frac{d\Phi'}{dr}
-4\pi G\rho'+\frac{l(l+1)}{r^2}\Phi'.
\end{eqnarray}
Using equations (\ref{acous_eq_1}-\ref{acous_eq_4}), the derivative of equation 
(\ref{acous_eq_2}) gives
\begin{equation}
\label{acous_single_eq}
\frac{d^2p'}{dr^2} + \frac{2}{r} \frac{dp'}{dr}
+\left[ \frac{\omega^2+4\pi G\rho_0}{c_0^2} - \frac{l(l+1)}{r^2} \right] p'=0.
\end {equation}
When searching for a spherical solution $p'=P'/r$, $P'$ must verify
\begin{equation}
\label{acous_final_eq}
\frac{d^2P'}{dr^2}
+\left[ \frac{\omega^2+4\pi G\rho_0}{c_0^2} - \frac{l(l+1)}{r^2} \right] P'=0.
\end {equation}
This typical wave equation with slowly variable coefficients has a well known 
solution in the JWKB approach (see appendix \ref{The JWKB Approach}), so that
\begin{equation}
\label{p'_sol}
p'(r)=\frac{A}{r\sqrt{k_r}} \cos \left(- \int_{r_0}^{r}\!\! k_r dr + \psi 
\right).
\end {equation}
By searching a solution of the same type for $\Phi'$, equation 
(\ref{acous_eq_4}) gives the particular solution
\begin{equation}
\label{Phi'_sol}
\Phi'(r)=\frac{4\pi G}{c_0^2k^2} \frac{A}{r\sqrt{k_r}} \cos \left(- 
\int_{r_0}^{r}\!\! k_r dr + \psi \right)
\end {equation}
and finally equation (\ref{acous_eq_2}) gives
\begin{equation}
\label{ksir_sol}
\xi_r(r)=\frac{A \sqrt{k_r}}{r \rho_0 c_0^2k^2} \sin \left(- \int_{r_0}^{r}\!\! 
k_r dr + \psi \right).
\end {equation}
Thus $\xi_r$ differs by $\pi /2$ in phase with $p'$, $\rho'$ and $\Phi'$. Remark 
also that $r_0$ and $r$ should be in a domain where $k_r \geq 0$. Otherwise 
exponential solutions should be considered. To be exhaustive, a general solution 
has to be added to the particular solution $\Phi'$, but it will play a role only 
for radial modes very near the center, so we will invoke it only in section 
\ref{Internal Turning Point leq 0}.

Equations (\ref{p'_sol}-\ref{ksir_sol}) are the expressions of a spherical wave 
that is well defined only locally. The amplitude $A$ and the phase $\psi$ are 
not constant but almost constant. At a given radius $r$, the radial wavenumber 
$k_r$ is given by
\begin{equation}
\label{kr2}
k_r^2=\frac{\omega^2+4\pi G\rho_0}{c_0^2} - \frac{l(l+1)}{r^2}
\end {equation}
and, for a spherical wave, its horizontal component is
\begin{equation}
\label{kh2}
k_h^2=\frac{l(l+1)}{r^2}.
\end {equation}

Thus the wave number is 
\begin{equation}
\label{k2}
k^2=\frac{\omega^2+4\pi G\rho_0}{c_0^2}.
\end {equation}
The amplitude of the radial displacement is further evaluated in appendix 
\ref{Eigenfunction}. The sound speed, or the phase velocity of the sound wave is 
(in a propagative medium where $k^2 >0$)
\begin{equation}
\label{phase_speed}
v_\varphi=\frac{\omega}{k}=\frac{c_0}{\sqrt{1+4\pi G\rho_0 /\omega^2}}
\end {equation}
which is equal to $c_0$ only when $4\pi G\rho_0/\omega^2$ is negligible compared 
to 1.

This propagative wave solution is  surprisingly  simple. Discussions about its 
relevance are needed, as well in the interior than at the surface of the star.

\subsection{Relevance in the Stellar Interior}
\label{Relevance in the Stellar Interior}
In the stellar interior, a wave solution given by equations 
(\ref{p'_sol}-\ref{ksir_sol}) are all the more simple as the gravitational 
perturbation is already taken into account. There is no need to develop to 
higher orders in $\Phi'$. The latter is considered at the same rank than the 
other perturbations $\xi$, $p'$, and $\rho'$, like it should be naturally. The 
final results obtained should be more precise and more realistic. Indeed, as 
only the relative gradients of equilibrium quantities are neglected, and not the 
gradients themselves, the gravity that is linked to the pressure gradient is not 
neglected. Physically, acoustic waves have enough short wavelengths not to see 
the gravitational stratification at large scale, but they will progressively 
pass through a non constant, hydrostatic structure governed by gravity. The 
non-negligible gravitational potential and its perturbation are seen by the term 
$4\pi g\rho'$ in equation (\ref{acous_eq_4}).

The consequence is that the obtained accuracy will be the same for all ranges of 
$l$ or $\omega$ (except for frequencies near gravity frequencies), contrary to 
restrictions of large $l$ or $\omega$ imposed by more drastic asymptotic 
approximations.

Another advantage is the autocoherence concerning the spherical eigenvalue 
$l(l+1)$ which will be maintained the same all along the present study. There is 
no need to replace it by $(l+0.5)^2$, a trick sometimes employed at low-$l$ (!), 
whose the artificial character is clearly pointed out in \citet{Langer_1937}. It 
is also easy now to bring an answer to the controversial question about the fact 
that the radial mode ($l=0$) reaches or not the center: due to its spherical 
nature, described by the term in $A/r$ in equation (\ref{p'_sol}), the wave can 
never touch the center, whatever the value of $l$. The non-radial waves reach 
their internal turning points where $k_r =0$, while the case of radial waves are 
examined in details in section \ref{Internal Turning Point leq 0} below.

\subsection{Relevance at the stellar Surface}
\label{Relevance at the stellar Surface}
At the surface, the term $4\pi G\rho_0$ can be neglected, equation (\ref{kr2}) 
becomes
\begin{equation}
\label{kr_surf}
k_r^2=\frac{\omega^2}{c_0^2} - \frac{l(l+1)}{r^2}
\end {equation}
and generally, for not too large $l$,
\begin{equation}
\label{kr_surf_lowl}
k_r=\frac{\omega}{c}
\end {equation}
which is particularly simple. The important issue is that $k_r$ continuously 
increases toward the surface. Yet it was established by current asymptotic 
approximations that the expression of $k_r$ goes to zero near the surface, and 
that location could be considered as the external turning point. Does it mean 
that equation (\ref{kr_surf}) is obtained in a too approximated context, and 
some terms could have been missed? It is actually possible to derive a less 
approximated propagation equation than equation (\ref{acous_final_eq}), in the 
following way.

At the surface, and for not too large $l$, it is legitimate to remove the terms 
$2/r$, $l(l+1)/r^2$, and the terms in $\Phi'$ from equations 
(\ref{init_eq_1}-\ref{init_eq_2}), which become
\begin{eqnarray}
\label{gen_asymp_1}
\frac{d\xi_r}{dr}&=& \frac{1}{\Gamma_1 H_p}\xi_r
    - \frac{p'}{\rho_0 c_0^2} \\
\label{gen_asymp_2}
\frac{dp'}{dr}&=& \rho_0(\omega^2-N^2)\xi_r - \frac{1}{\Gamma_1 H_p}p'. 
\end{eqnarray}
Providing that
\begin{equation}
\label{d_con2_cson2}
\frac{1}{c_0^2} \frac{dc_0^2}{dr} = 
       \frac{1}{\Gamma_1 p_0} \frac{d(\Gamma_1 p_0)}{dr} + \frac{1}{H_\rho},
\end{equation}
\begin{equation}
\label{N2_cson2}
\frac{N^2}{c_0^2} = -\frac{1}{(\Gamma_1 Hp)^2} + \frac{1}{\Gamma_1 H_p H_\rho}
\end{equation}
and using equations (\ref{gen_asymp_1}-\ref{gen_asymp_2}) in the derivation of 
equation (\ref{gen_asymp_1}), one obtains:
\begin{eqnarray}
\label{gen_asymp_ksir}
\!\!\!\!\!\!\!\!
\frac{d^2\xi_r}{dr^2} \!\!\!\! &+& \!\!\!\! \frac{1}{\Gamma_1 p_0}
\frac{d(\Gamma_1 p_0)}{dr} \frac{d\xi_r}{dr} \nonumber \\
\!\!\!\! &+& \!\!\!\!
 \left( \frac{\omega^2}{c_0^2}-\frac{1}{\Gamma_1 H_p H_\rho} + 
\frac{1}{\Gamma_1 p_0} \frac{d^2 p_0}{dr^2} \right) \xi_r =0
\end{eqnarray}
To avoid the term in $d\xi_r/dr$, introduce $K$ as:
\begin{equation}
\label{ksir_K}
\xi_r=K \exp\left(-\frac{1}{2}\int_{}^{}\!\! \frac{1}{\Gamma_1 p_0} 
\frac{d(\Gamma_1 p_0)}{dr} \! dr \right)
\end{equation}
to obtain the propagation equation:
\begin{eqnarray}
\label{gen_asymp_K}
&& \!\!\!\!\!\!\!\!\!\!\!\!
\frac{d^2K}{dr^2} + \Bigg\{
\frac{\omega^2}{c_0^2} - \frac{1}{\Gamma_1H_pH_\rho} 
+ \frac{1}{\Gamma_1 p_0} \frac{d^2p_0}{dr^2}
 \\
&& \!\!\!\!\!\!\!\!\!\!\!\!
-\frac{1}{4} \left[ \frac{1}{\Gamma_1 p_0} \frac{d(\Gamma_1 p_0)}{dr}\right]^2
-\frac{1}{2} \frac{d}{dr} 
\left[ \frac{1}{\Gamma_1 p_0} \frac{d(\Gamma_1 p_0)}{dr} \right]
\Bigg\} K =0. \nonumber
\end{eqnarray}
Equation (\ref{gen_asymp_K}) is a generalization of what is derived in current 
asymptotic approximations. Indeed, when $g$ is constant, the two terms after 
$\omega^2/c_0^2$ cancel each other out, and if further more $\Gamma_1$ is 
constant, it looks more familiar \citep{Deubner&Gough_1984}:
\begin{equation}
\label{gen_asymp_K_appr}
\frac{d^2K}{dr^2} + \left[ \frac{\omega^2}{c_0^2} 
- \frac{1}{4H_p^2}\left( 1+2\frac{dH_p}{dr}\right) \right] K=0.
\end{equation}
At the surface, these last equations are indisputably less approximated than 
those obtained in the local homogeneity approximation. The deduced $k_r$ goes to 
zero near the surface. But for all that, are those last expressions more 
relevant? Can them be used to determine the external turning points?

To answer, a detour by the JWKB approach is necessary. Following appendix 
\ref{The JWKB Approach}, equations (\ref{gen_asymp_K}, \ref{gen_asymp_K_appr}) 
admit a wave-like solution only when the conditions of equation 
(\ref{JWKB_cond}) are verified, i.e. when properties of the medium vary slowly. 
The JWKB approximation also predicts that in these conditions, there is no wave 
reflection, or at least the reflections are negligible. In the opposite case, 
there is no wave-like solution and reflections will occur.

An important consequence is that, generally, in order to be able to apply JWKB, 
the expression of $k_r(r)$ can only contain medium variables at the same 
derivation order. For example, if the equilibrium pressure $p_0$ appears in 
$k_r(r)$, then the derivatives $dp_0/dr, d^2 p_0/dr^2$ must be negligible, if 
not the conditions of equation (\ref{JWKB_cond}) cannot be verified (except for 
very specific cases). If it contains $dp_0/dr$, then it should contain neither 
$p_0$ nor $d^2p_0/dr^2,d^3p_0/dr^3$.

Another consequence is that when the JWKB approximation is applicable, and only 
in this case, the sign of $k_r^2$ indicates the propagation or evanescent 
character. It is also clear that when $k_r(r)=0$, the conditions of equation 
(\ref{JWKB_cond}) are of course violated, and there is reflection. But we should 
not forget that reflection immediately occurs whenever those conditions are no 
more valid, even when $k_r(r)$ is very far from zero. And we should keep in mind 
that \emph{those reflection conditions are not contained in the expression of 
$k_r$ alone}. 

It appears therefore that the $k_r$ deduced from equations (\ref{gen_asymp_K}) 
or (\ref{gen_asymp_K_appr}) contains $p_0, \rho_0$ as well as their derivatives, 
thus is generally rapidly variable, thus the JWKB approach cannot be applied, 
and its zeros have no special physical sense. It just means that when the 
relative derivatives of equilibrium quantities cannot be neglected, we are in 
the evanescent part, outside the resonant cavity. This is coherent with the 
local-homogeneity analysis. (In fact, equation (\ref{gen_asymp_K_appr}) 
indicates that the JWKB method can nevertheless be applied in the very special 
case where both $c_0$ and $H_p$ are slowly variable, so that $H_p^{-2} dH_p/dr$ 
can be neglected. It is the case for an isothermal atmosphere which will be 
discussed further in section \ref{Discussion on the Outer Boundary Conditions}).

\begin{figure}[hbt]
\centering
\plotone{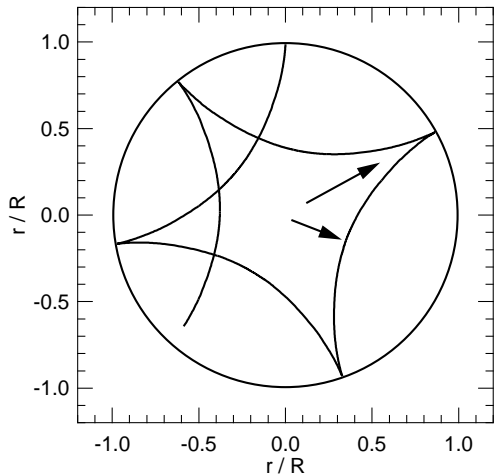}
\caption{Wave trajectories for the mode $l=5, n=5$. Internal and external 
turning points are indicated by the arrows, which show their different 
behaviours.}
\label{fig_trajectories}
\end{figure}

On the contrary, in the context of local homogeneity, i.e. slowly variable 
medium, the $k_r$ obtained in equation (\ref{kr2}) \emph{automatically} 
satisfies the JWKB application conditions. In the present case of spherical 
modes, the internal cavity limits are generally situated towards the center 
where the medium is really slowly variable, and the turning points will be 
determined by $k_r(r)=0$, or very close to that. Concerning the external limit, 
the cause of rapidly variable medium should rather be invoked. In other words, 
an acoustic wave turns back in the interior because of its own spherical nature, 
and at the surface because of an abrupt medium change. This internal/external 
dissymmetry can be seen on the wave trajectories (Fig. \ref{fig_trajectories}).

From that long but necessary discussion, it can be concluded that conceptually, 
the local homogeneity approximation and its cavity limit criteria are nothing 
but a physical translation of the JWKB mathematical expression.

%
\subsection{Wave Trapping}
\label{Wave Trapping}
When the acoustic wave is reflected at the external and internal limits of a 
star, we are in the presence of standing waves inside a cavity, where only 
determined frequencies will be favoured. The sum of forward and backward waves 
forms standing waves with nodes and antinodes separated by $\pi$ in phase
\begin{eqnarray}
y(r)&=&\frac{A}{r}\cos\left( \omega t - \int_{r_0}^{r}\!\! k_r dr + \psi \right)
\nonumber \\
&+&\frac{A}{r}\cos\left( \omega t - \int_{r}^{r_0}\!\! k_r dr + \psi' \right) 
\nonumber \\ \nonumber
\end{eqnarray}
\begin{equation}  \label{standing_wave}
= \frac{2A}{r}\cos(\omega t+\frac{\psi+\psi'}{2}) \cos\left(-\int_{r_0}^{r}\!\! 
k_r dr +\frac{\psi-\psi'}{2} \right)
\end{equation}
where $y$ stands for $\xi_r$, or $p'$, or $\rho'$ or $\Phi'$ . Following Figure 
\ref{fig_standing_wave}, it is clear that the phase advance between the turning 
points $r_1$ and $r_2$ should verify
\begin{equation} \label{Duvall}
\int_{r_1}^{r_2}\!\!k_r dr=(n+\alpha)\pi
\end{equation}
where $n$ is the number of nodes along r, called the mode order, and $\alpha$ is 
a parameter to adjust following the phase conditions at $r_1$ and $r_2$. For 
example, if there are two antinodes at the two limits, $\alpha=0$, if there are 
one node and one antinode, $\alpha=-0.5$. If there are two nodes, $\alpha=-1$. 
The convention adopted here is to count every node of the \emph{displacement} 
$\xi_r$.
\begin{figure}[htb]
\centering
\plotone {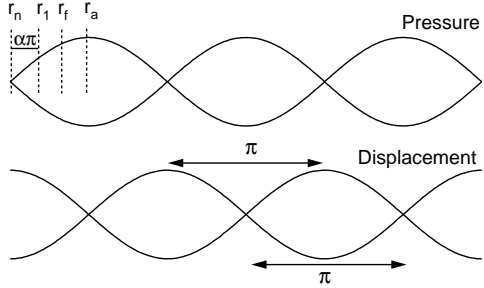}
\caption{Acoustic standing wave with nodes and antinodes separated by $\pi$ in 
phase. Pressure and displacement differs by $\pi/2$ in phase. For the pressure, 
the locations of the internal node, turning point, fitting point, and antinode 
are marked by $r_n$, $r_1$, $r_f$, and $r_a$}.
\label{fig_standing_wave}
\end{figure}
%

The relation (\ref{Duvall}) is largely used in different approaches for 
interpreting solar acoustic eigenfrequencies \citep{Duvall_1982}. Even so, 
attention must be paid that $r_2$ is sometimes assimilated to R, the star 
radius, while it is \emph{not} the case here. This apparently small subtility is 
in fact fundamental. Indeed, assuming $r_2\equiv R$ is a crude approximation 
leading to frequency errors of tens of $\mu$Hz. Any attempt to compensate it, 
amounts to make a variable change while integrating (\ref{Duvall}), therefore to 
work with a wavenumber $k_r$ that is not the real one. So $\alpha$ is not the 
same than that employed here. Moreover, the phase conditions at external and 
internal turning points are completely mixed. The overall consequence is that 
one is not working in the real world, but in a kind of phase space.

Here, by working with $r_2$ and not R, the eigenmode determination becomes 
similar to what is done for an acoustic cavity on Earth, just like for any 
musical instrument. The eigenfunctions projected on the radial direction are 
given by the standing wave expression (eq. \ref{standing_wave}), and the 
eigenfrequencies are determined by the Duvall expression (\ref{Duvall}). Thus, 
to obtain the full solution it only remains to determine the internal and 
external turning points together with the phase conditions at these points, i.e. 
namely $r_1$, $r_2$ and $\alpha$.

\subsection{External Turning Point}
\label{External Turning Point}
Let us begin to look at the external reflection conditions because they are the 
simplest ones, for the phase as well as for the location.

Following what has been discussed in section \ref{The Present Approach}, at the 
surface, the acoustic wave is reflected when a `pressure vacuum' is encountered. 
For the wave, everything looks like the end of the matter. Like any matter wave 
encountering a `matter vacuum', there is an antinode at this point for the 
displacement (think e.g. about longitudinal standing waves along a spring with a 
free end). That determines the phase.

\begin{figure}[htb]
\centering
\plotone {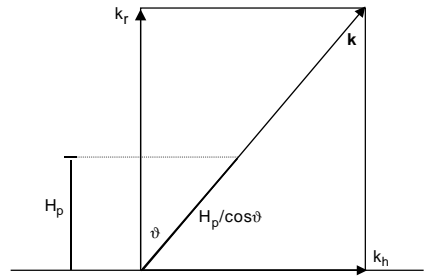}
\caption{Projection of $H_p$ on the propagation direction $ \mathbf{k} $, at the 
surface. The surface curvature of the Sun has been neglected.
        }
\label{fig_Hp_k}
\end{figure}

\begin{figure}[htb]
\centering
\plotone {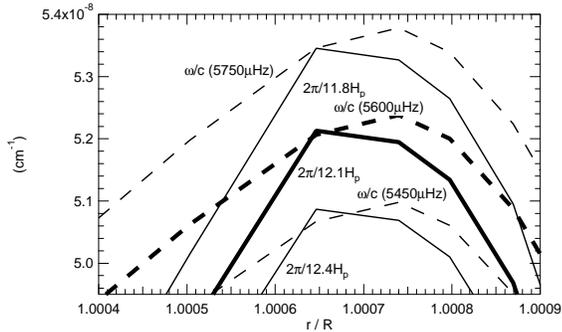}
\caption{Determination of the coefficient $a_v$ in equation 
(\ref{prop_cond_ext_asymp}). Continuous curves: $2\pi /a_v H_p$ following the 
gaz pressure of the model C of \citet{VAL_1981}. Dashed curves: $\omega /c$. 
When $a_v=12.1$, the mode at 5600 $\mu$Hz is the last one trapped.
        }
\label{fig_ae}
\end{figure}

For the location of reflection, we must look at the projection of $H_p$ onto the 
propagation direction. Neglecting the surface curvature (see Figure 
\ref{fig_Hp_k}), the propagation condition, equation (\ref{prop_cond}), becomes
\begin{equation} \label{prop_cond_ext}
\frac{k^2}{k_r}=\frac{1}{k_r}\frac{\omega^2}{c_0^2}>\frac{2\pi}{a_v H_p}
\end{equation}
knowing that the phase speed (eq. \ref{phase_speed}) at the surface is $c_0$. 
The constant $a_v$, v standing for vacuum, remains to be determined. This task 
concerns typically the boundary conditions that \emph{cannot} be already 
included in the initial differential equations. Something else than the physics 
of wave propagation should be considered. In order to avoid theoretical results 
which are not totally reliable at the surface,  we can choose the experimental 
Sun's cut-off frequency combined with the semi-empirical atmosphere of 
\citet{VAL_1981}. According to \citet{Fossat_etal_1992} and 
\citet{Garcia_etal_1998}, the observed Sun's cut-off frequency is
\begin{equation} \label{cut-off_frequency}
5600  \pm 150 \mu Hz.
\end{equation}
In this frequency range and at the surface, equation (\ref{prop_cond_ext}) can 
be simplified to
\begin{equation} \label{prop_cond_ext_asymp}
\frac{\omega}{c_0}>\frac{2\pi}{a_v H_p}.
\end{equation}
Figure \ref{fig_ae} shows that, in order that the last mode trapped is at 5600 
$\mu$Hz, $a_v$ should be:
\begin{equation} \label{av}
a_v=12.1 \pm 0.3
\end{equation}
This value could be refined by more sophisticated methods. For the moment, note 
that a change of $a_v$ by +0.1, will lead to a general frequency shift of about 
-2 $\mu$Hz.


%

%
\begin{figure}[htb]
\centering
\plotone {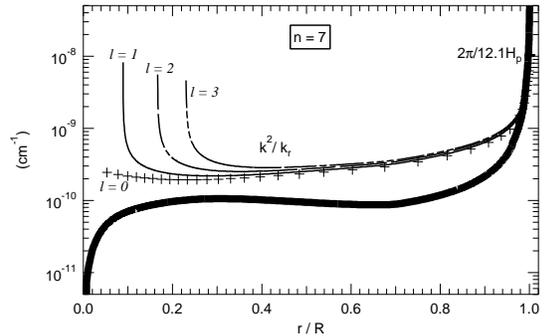}
\caption{Some examples of external turning points determined by the intersection 
of the two curves expressed in equation (\ref{prop_cond_ext}): 
$k^2/k_r$ for $n=7$ and $l=0$, 1, 2, 3 (symbols described in 
fig.\ref{fig_legende} ), and $2\pi$/$12.1H_p$ (thick continuous line).
       }
\label{fig_ex_r2}
\end{figure}


%
Conceptually speaking, the determination of the cavity external limit of a star 
is similar to that of a wind musical instrument with an open end. There, like 
for the Sun, one cavity end is not materially delimited while the 
eigenfrequencies are precisely defined. There also, the absence of an elaborate 
theory can be overcome by the use of experimental results to determine the 
equivalent external limits \citep[p.~201-202]{Rayleigh_1896}.

So all the external reflection conditions are defined.

In Figure \ref{fig_ex_r2} are given some examples of external turning points 
determined by the intersection of the two curves expressed in equation 
(\ref{prop_cond_ext}), for $l=0$, 1, 2, 3 and $n=7$. In such a diagram, the very 
low frequency modes  $l=0$, 1 and $n=1$, 2 will have their representative curves 
under that of $2\pi/12.1H_p$ at several locations, which means that they hardly 
exist, the Sun structure being too inhomogeneous from their wavelength point of 
view.

\subsection{Internal Turning Point ($l\neq 0$)}
\label{Internal Turning Point lneq 0}

In the deep interior where the structure can be considered as locally 
homogeneous, there is no problem of boundary conditions  and thus no genuine 
reflection point. (Except for some very low $l, n$ modes, see the section above, 
and for the $l=0$ modes, see the next section). Nevertheless, the spherical 
symmetry implies that there always exists an innermost point, called the 
internal turning point $r_1$, the location where the radial component of the 
wave vector vanishes. And, when studying only the radial part, we need to 
determine this internal turning point and the phase condition there. For 
non-radial modes, i.e. with $l$ and $k_h\neq 0$, $r_1$ is given by
\begin{equation} \label{r1cond_lneq0}
k_r=0 \Leftrightarrow \frac{c_0^2}{r_1^2}-\frac{4\pi 
G\rho_0}{l(l+1)}=\frac{\omega^2}{l(l+1)}.
\end{equation}

When the gravitational contribution is neglected ($G\rho_0=0$), the expression 
(\ref{r1cond_lneq0}) is reduced to the well known one used in the context of 
current asymptotic approximations
\begin{equation} \label{r1cond_lneq0_asymp}
\frac{c_0^2}{r_1^2}=\frac{\omega^2}{l(l+1)}.
\end{equation}

\begin{figure}[htb]
\centering
\plotone {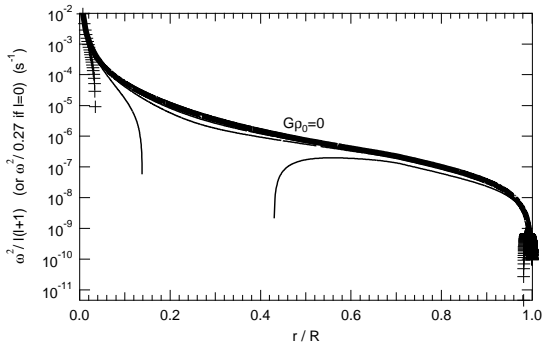}
\caption{Right or left member of equation (\ref{r1cond_lneq0}) and 
(\ref{r1cond_leq0}), as a function of radius. The cases of $l=0$, 1, 2, 3, in 
the presence of gravity (symbols as defined in Figure \ref{fig_legende}) are 
compared to the case without gravity (upper thick continuous line).
        }
\label{fig_r1cond}
\end{figure}

Some turning points can already be seen on Figure \ref{fig_ex_r2} where 
$1/k_r\rightarrow\infty$. It is also interesting to represent the right or left 
member of equation (\ref{r1cond_lneq0}) as a function of the radius (Figure 
\ref{fig_r1cond} for the Sun), so that $r_1$ can be immediately determined once 
the frequency and the degree of a mode are known. For a given $\omega$, the 
larger the degree, the less the mode goes deep. For a given degree, the larger 
the $\omega$, the deeper the mode goes. For the degrees $l\geq 5$, the curves 
begin to merge together with that where $G\rho_0=0$, which is independent of 
$l$. For the smaller degrees, the difference with the case without gravity is 
appreciable, even far from the centre, until the middle of the solar radius, 
because while $\rho_0$ decreases with the radius, $c_0^2/r^2$ decreases also 
quickly. The curve for the $l=1$ modes present a further special feature: the 
gravitation term becomes much more important compared to $c_0^2/r^2$, so that a 
gap appears in the radiative zone, separating the curve into 2 parts, and for 
the concerned frequencies, the outermost turning point $r_1/R$ is drastically 
reduced to 0.137 instead of 1 for the other degrees.


%

The turning points $r_1/R$ are given in Table \ref{table_r1} for different $l$ 
values at typical levels of $\omega^2/l(l+1)$, to be compared to the case 
without gravity to appreciate the contribution of gravity for the Sun.
\begin{table}[htb]
\caption{Internal turning points $r_1/R$ in the presence of gravity, for 
different $l$ at typical levels of $\omega^2/l(l+1)$, compared to the case 
without gravity.}
\label{table_r1}
$$ \begin{array}{ccccc}
 \hline
 &\frac{\omega^2}{l(l+1)}= & 10^{-5}  & \ \ \ 10^{-6}\ \ \  & 3\,10^{-7} \\
 \hline
 l=1                     & & 0.110    & 0.134    & 0.137       \\
 l=2                     & & 0.168    & 0.388    & 0.621       \\
 l=5                     & & 0.198    & 0.445    & 0.646       \\
 l=10                    & & 0.203    & 0.452    & 0.650       \\
 \hline
Without \, G             & & 0.205    & 0.454    & 0.652       \\
\end{array} $$
\end{table}

Globally, when the gravitational potential is taken into account, $r_1$ is 
smaller, especially for $l=1$. Gravity pulls the low $l$ modes toward the 
centre.

The location of the internal turning point can be directly determined by a 
mathematical expression, while the phase condition at this point is less easy to 
find out. A discussion about $H_p$ like at the surface is of no use because here 
its projection on the propagation direction vanishes. Let us come back to 
equation (\ref{acous_final_eq}) governing $P'=rp'$, which admits the JWKB 
standing-wave solution
\begin{equation}
\label{P'_standing_sol}
P'(r)=\frac{2A}{\sqrt{k_r}} \cos \left(- \int_{r_0}^{r}\!\! k_r dr + 
\frac{\psi-\psi'}{2} \right).
\end {equation}
This solution fails of course near $r=r_1$ where $k_r=0$. A well known method to 
bypass this difficulty is to develop linearly $k_r^2$ around $r_1$
\begin{equation}
\label{kr2lin}
k_r^2=k_1(r-r_1)
\end {equation}
and to adopt a new variable $x$
\begin{equation}
\label{x_definition}
x=k_1^{1/3} (r-r_1)
\end {equation}
where $k_1$ is the derivative of $k_r^2$ in $r_1$. Equation 
(\ref{acous_final_eq}) becomes
\begin{equation}
\label{P'(x)_eq}
\frac{d^2 P'}{dx^2}+xP'=0
\end {equation}
which admits the general solution
\begin{equation}
\label{P'(x)_airy_sol}
P'=C\mbox{Ai}(-x)+D\mbox{Bi}(-x)
\end {equation}
where $C$ and $D$ are the constants to be determined, Ai and Bi the Airy 
functions which present an undulatory behaviour for positive x and an 
exponential behaviour for negative x. As the Ai function is exponentially 
decreasing toward the center while the Bi function is increasing, one must 
choose $D=0$. The next step is to fit the two solutions (\ref{P'_standing_sol}) 
and (\ref{P'(x)_airy_sol}) at a fit location $r_f$. This latter is usually 
chosen so that the corresponding $x_f$ is very large, in order to use the 
asymptotic development of Ai which is sinusoidal, and to deduce easily the wave 
phase. In fact, this method introduces an important lack of precision because 
the Airy solution is only valid when (\ref{kr2lin}) is valid, that is around 
$x=0$. We will here adopt fitting points $r_f$ that are a priori not far from 
$r_1$, corresponding to small values of $x_f$. Consider now the standing 
solution coming from the nearest antinode to $r_1$, located at $r_a$, we have 
$(\psi-\psi')/2=0$ in equation (\ref{P'_standing_sol}). When matching this 
solution and its derivative to the Airy solution at the the fitting point:
\begin{equation}
\label{fitting_P}
\frac{2A}{\sqrt{k_r}} \cos \int_{r_a}^{r_f}\!\!\!\! k_r dr=C\mbox{Ai}(-x_f) 
\end{equation}
\begin{eqnarray}
\label{fitting_P'}
2A\sqrt{k_r} \sin \int_{r_a}^{r_f}\!\!\!\! k_r dr
-Ak_r^{-3/2} \frac{dk_r}{dr}\cos\int_{r_a}^{r_f}\!\!\!\! k_r dr \nonumber \\ 
\!\!\!\!\!\!\!\!\!\!\! =-C\mbox{Ai'}(-x_f)k_1^{1/3}
\end{eqnarray}
the phase of the standing wave is imposed :
\begin{equation}
\label{intkr_ra_rf}
\int_{r_a}^{r_f}\!\!\!\! k_r dr=-\arctan \left[ 
\frac{1}{4k_r^3}\frac{dk_r^2}{dr} - 
\frac{k_1^{1/3}}{k_r}\frac{\mbox{Ai'}(-x_f)}{\mbox{Ai}(-x_f)}
\right]
\end{equation}
where Ai' stands for the derivative of Ai over $x$. As $\alpha$ is the phase lag 
between $r_1$ and a displacement antinode, i.e. a pressure node located in 
$r_n$, it is given by (see fig. \ref{fig_standing_wave})
\begin{equation}
\label{alpha_pi}
\alpha \pi=-\int_{r_n}^{r_1}=-\int_{r_n}^{r_a}-\int_{r_a}^{r_f}-\int_{r_f}^{r_1}
\end{equation}
where the terms $k_r dr$ inside the integral signs are implicit. This matching 
has only a sense when equation (\ref{kr2lin}) can be used. Applying this latter 
and equation (\ref{x_definition}), then calculating the first and third 
integrals of equation (\ref{alpha_pi}), the final expression of $\alpha$ can be 
expressed as a function of $x_f$:
\begin{equation}
\label{alpha_pi_xf}
\alpha\pi=-\frac{\pi}{2}+\arctan \left\{ \frac{1}{x_f^{0.5}} 
\left[\frac{1}{4x_f}-\frac{\mbox{Ai'}(-x_f)}{\mbox{Ai}(-x_f)} \right] 
\right\}+\frac{2}{3}x_f^{3/2}.
\end{equation}

It remains now to determine the fitting point. It should be chosen at the 
farthest limit toward $r_1$ where the JWKB solution is still valid, to be also 
in the validation domain of the Airy solution. As seen in appendix \ref{The JWKB 
Approach}, the JWKB method is valid as long as the derivative of $k_r$ is enough 
small compared to $k_r$:
\begin{equation}
\label{JWKB_cond_principe}
\frac{dk_r}{dr} \leq ak_r^2 \Leftrightarrow 
\frac{1}{2k_r^3} \frac{dk_r^2}{dr} \leq a
\end{equation}
where a is a `small' number. In other words, the variation of $P'$ must be slow 
enough on a wavelength when $k_r$ goes to zero. This condition should be 
coherent with that at the external limit which is of the same nature (eq. 
\ref{prop_cond}), except that the characteristic scale height $H_f$ to be 
considered here does not come from the environment but from the function $P'$ 
itself. We can define a function f
\begin{equation}
\label{function_f}
f=\frac{1}{\sqrt{k_r}}
\end{equation}
and should use the already determined coefficient $a_v$ to write the condition
\begin{equation}
\label{JWKB_cond_Hf}
\frac{2\pi}{k_r} \leq a_v H_f
\end{equation}
with
\begin{equation}
\label{Hf}
H_f^{-1}= -\frac{1}{f} \frac{df}{dr}=\frac{1}{4k_r^2} \frac{dk_r^2}{dr}.
\end{equation}
The searched condition (\ref{JWKB_cond_principe}) is thus
\begin{equation}
\label{JWKB_cond_final}
\frac{1}{2k_r^3} \frac{dk_r^2}{dr} \leq \frac{a_v}{\pi}.
\end{equation}
Using equation (\ref{kr2lin}), a first fitting point can then be obtained
\begin{equation}
\label{rf1}
r_{f1}=r_1+ \frac {(\pi/2a_v)^{2/3}} {k_1^{1/3}}.
\end{equation}
Imposing that the ratio $k_1/k_r^3$ is enough small, the condition 
(\ref{JWKB_cond_final})  is in fact not sufficient to warranty the validity of 
the standing wave solution. When $l$ and $n$ are small, the slope $k_1$ of the 
function $k_r^2$ can be so close to zero that $k_r^2$ will also remains very 
close to zero. The wave solution loose its meaning because the wavelength 
becomes too large. Either $k_1$ or $k_r^2$, or their product should be enough 
large. The smallest $k_r$ is determined by the largest wavelength that can be 
contained in the cavity:
\begin{equation}
\label{kr_limit}
\lambda_r \leq r_2-r_1 \Rightarrow k_r \geq \frac{2\pi}{r_2-r_1}.
\end{equation}
The smallest slope $k_1$ should be so that $k_r$ reaches an `appreciable' value 
, let us say corresponding to $\lambda_r\leq r_2$, on a `negligible' distance, 
i.e. a distance $(r-r_1)$ small enough compared to the wavelength. To be 
coherent with above, we use again the coefficient $a_v$ and the relation 
(\ref{prop_cond}) with the inverse inequality sign
\begin{equation}
\label{negligible_distance}
\lambda=\frac{2\pi}{k_r} \geq a_v (r-r_1).
\end{equation}
That leads to
\begin{equation}
\label{k1_limit}
k_1 \geq a_v\frac{(2\pi)^2}{r_2^3}
\end{equation}
which, together with equation (\ref{kr_limit}) allow to arrive to the searched 
condition on the product $k_1k_r^2$
\begin{equation}
\label{kr2k1_limit}
k_1 k_r^2 \geq a_v\frac{(2\pi)^4}{(r_2-r_1)^2r_2^3}.
\end{equation}
Using equation (\ref{kr2lin}), the second fitting point can be obtained
\begin{equation}
\label{rf2}
r_{f2}=r_1+\frac{a_v(2\pi)^4}{(r_2-r_1)^2 r_2^3 k_1^2}.
\end{equation}
The final fitting point can be taken as a compromise between the conditions 
(\ref{JWKB_cond_final}) and (\ref{kr2k1_limit}), that is a mean value between 
$r_{f1}$ and $r_{f2}$
\begin{equation}
\label{rf}
r_f=r_1+\frac {(\pi/2a_v)^{2/3}} {2 k_1^{1/3}}+\frac{a_v(2\pi)^4}{2(r_2-r_1)^2 
r_2^3 k_1^2}.
\end{equation}

Everything is now ready for the complete calculation of the $l\neq0$ modes.

\subsection {Internal Turning Point ($l=0$)}
\label{Internal Turning Point leq 0}

For the radial modes
\begin{equation}
 l=0, \ k_h=0, \ k_r=k>0, 
\end{equation}
there is \emph{a priori} no turning point as for $l\neq 0$. It is very tempting 
to conclude that these modes pass through the star center which is a node, since 
due to symmetrical reasons it must stay immobile. In this case:
\begin{equation} 
r_1=0, \ and \ \alpha=-0.5.
\end{equation}
But this induces eigenfrequencies that are lower than numerical or observed 
results by several tens of $\mu$Hz. In fact, no mode can touch the center: the 
linear solution of local wave cannot be applied at $r=0$. For a spherical wave, 
it is obvious that the center is a singular point according to its expression 
(\ref{standing_wave}). The wave must be stopped somewhere before reaching the 
center. As the perturbed pressure $p'$ goes to infinity, for the wave, 
everything happens as if the surrounding pressure goes to zero. And everything 
looks like the vacuum condition at the external limit, except that here the 
decreasing `external' pressure is directly given by a law in $1/r$. Thus 
$H_p=r$, and the condition (\ref{prop_cond}) gives the turning point $r_1$
\begin{equation} 
\label{r1cond_leq0}
k_r=\frac{2\pi}{a_v r}.
\end{equation}
Taking into account the numerical value of $a_v$ (Eq. \ref{av}), this relation 
can be written as
\begin{equation} 
\label{r1_l=0_num}
\frac{\omega^2+4\pi G\rho_0}{c_0^2}-\frac{(0.51927)^2}{r_1^2}=0.
\end{equation}
When neglecting the gravity, this formula is very close to that using by 
traditional approximations where the term $l(l+1)$ is artificially replaced by 
$(l+0.5)^2$. Therefore, the present concept of `vacuum condition' characterised 
by the coefficient $a_v$, could be considered as the justification of this usual 
way of doing for $l=0$.

Equation (\ref{r1_l=0_num}) can also be rewritten in a general form like 
Equation (\ref{r1cond_lneq0}):
\begin{equation} 
\label{r1cond_leq0}
\frac{c_0^2}{r_1^2}-\frac{4\pi G\rho_0}{(2\pi /a_v)^2}=\frac{\omega^2}{(2\pi 
/a_v)^2}
\end{equation}
This expression is also represented in Figure \ref{fig_r1cond} giving the 
internal turning point. The curve features an important gap in the radiative 
zone, even bigger than for the $l=1$ modes, and for the concerned frequencies 
the largest possible $r_1 /R$ can only go up to 0.035. 

The internal turning point being determined, let us examine the phase at this 
location. Whenever the gravity is neglected, we are in the same `vacuum' type 
condition than at the external boundary (Eq. \ref{vacuum_cond}) : the standing 
wave will present a pressure node or a displacement antinode. Thus, in this case
\begin{equation} 
\label{alpha=0}
\alpha=0.
\end{equation}

In the presence of a gravity field, the condition 
(\ref{vacuum_cond_under_gravity}) must be rather applied, which can be rewritten 
as
\begin{equation} 
\label{p1+rophi'=0}
p'(r_1)+\rho_0 \phi'(r_1)=0.
\end{equation}
As $r_1$ is very near the center, the exact solution of $\phi'$ must be used. 
Indeed, resulting from Poisson's equation, the gravitational potential as well 
as its perturbation must remain always finite, even at the center. Going back to 
equation (\ref{acous_eq_4}), and with $\phi'=F'/r$, the equation that has to be 
solved is
\begin{equation}
\label{F'_eq}
\frac{d^2F'}{dr^2}-\frac{l(l+1)}{r^2}\Phi'= -\frac{4\pi GP'}{c_0^2}.
\end{equation} 
To the particular solution given by equation (\ref{Phi'_sol}), must be added the 
homogeneous solution
\begin{equation}
\label{F'_sol-homogenous}
F'=Cr^{l+1}+\frac{D}{r^l}
\end{equation}
with C, D the constants to be determined. In order to have a negligible 
gravitational potential at large radius, $C=0$ must be chosen. And for $l=0$, D 
must be chosen so that the general solution is
\begin{equation}
\label{Phi'_sol_gen}
\Phi'(r)=\frac{4\pi G}{c_0^2k^2} \frac{A}{r\sqrt{k}} \left[ \cos \left(- 
\int_{0}^{r}\!\! kdr + \psi \right) -cos\psi \right].
\end {equation}
This keeps $\phi'$ finite at $r=0$. Equation (\ref{p1+rophi'=0}) becomes, for 
the corresponding standing waves
\begin{equation}
\left(\frac{c_0^2 k^2}{4\pi G\rho_0} +1 \right) \cos 
\left(\int_{0}^{r_1}\!\!\!\! kdr + \frac{\psi-\psi'}{2} \right)- \cos 
\left(\frac{\psi-\psi'}{2} \right)=0.
\end{equation}
Knowing that
\begin{equation}
\int_{0}^{r_n}\!\!\!\! -kdr + \frac{\psi-\psi'}{2} =\frac{\pi}{2}
\end{equation}
with $r_n$ the location of the closest node to $r_1$, and
\begin{equation}
\alpha \pi=-\int_{r_n}^{r_1}=\int_{0}^{r_n}-\int_{0}^{r_1}
\end{equation}
where the terms $kdr$ inside the integral signs are implicit, the equation 
determining $\alpha \pi$ can be deduced
\begin{equation}
\left(\frac{\omega^2}{4\pi G\rho_0} +2 \right) \sin \alpha \pi - \sin 
\left(\alpha \pi+ \int_{0}^{r_1}\!\!\!\! kdr \right)=0.
\end{equation}
As generally $\alpha \pi \ll 1$, this can be simplified to:
\begin{equation}
\alpha \pi=\frac{\sin\int_{0}^{r_1}\!\! kdr}{\omega^2 / 4\pi G\rho_0+2 
-\cos\int_{0}^{r_1}\!\! kdr}.
\end{equation}
As very near the center the wave number $k$ can be considered to be roughly 
constant,
\begin{equation}
\int_{0}^{r_1}\!\!\!\! kdr \simeq kr_1=\frac{2\pi}{a_v}
\end{equation}
a further simplified expression for $\alpha$ can be used
\begin{equation}
\label{alpha_pi_simplified}
\alpha \pi=\frac{\sin \frac{2\pi}{a_v}}{\frac{\omega^2}{4\pi G\rho_0}+2 
-\cos\frac{2\pi}{a_v}}.
\end{equation}
When the gravity potential is neglected ($G\rho_0 \rightarrow 0$), this 
expression is coherent with equation (\ref{alpha=0}).

\section{The Whole Set of Equations and Results}
\label{The Whole Set of Equations and Results}

We have now every ingredient to calculate the eigenfrequencies. It is worth 
gathering here the whole set of equations used for the general case in the 
presence of gravity. Given the 2 parameters $l$, $n$ which are respectively the 
mode degree and order, a system of 2 equations with 2 unknowns $\omega$ and 
$\alpha$ has to be solved:
\begin{equation}
\label{whole_set_Duvall}
\!\!\!\!\!\int_{r_1}^{r_2} \!\! k_r dr=(n+\alpha)\pi
\end{equation}
and for $l\neq 0$
\begin{equation}
\label{whole_set_alpha_lneq0}
\alpha\pi=-\frac{\pi}{2}+\arctan \left\{ \frac{1}{x_f^{0.5}} 
\left[\frac{1}{4x_f}-\frac{\mbox{Ai'}(-x_f)}{\mbox{Ai}(-x_f)} \right] 
\right\}+\frac{2}{3}x_f^{3/2}
\end{equation}
for $l=0$
\begin{equation}
\label{whole_set_alpha_leq0}
\alpha \pi=\frac{\sin \frac{2\pi}{a_v}}{\frac{\omega^2}{4\pi G\rho_0}+2 
-\cos\frac{2\pi}{a_v}}
\end{equation}
with
\begin{equation}
\label{whole_set_kr}
k_r=\sqrt{\frac{\omega^2+4\pi G\rho_0}{c_0^2}-\frac{l(l+1)}{r^2}}.
\end{equation}
$r_1$ is given by
\begin{equation}
\label{whole_set_r1}
\!\!\!\! k_r=0 \ \ for \ l\neq 0, \ \ \ k_r=\frac{2\pi}{a_v r} \ \ for \ l=0,
\end{equation}
and $r_2$ given by
\begin{equation}
\label{whole_set_r2}
\frac{k^2}{k_r}=\frac{2\pi}{a_v H_p}
\end{equation}
with
\begin{equation}
\label{whole_set_av}
a_v=12.1.
\end{equation}
$x_f$ is given by
\begin{equation}
\label{whole_set_xf}
x_f=k_1^{1/3}(r_f-r_1)
\end{equation}
where $k_1$ is the derivative of $k_r^2$ at $r_1$, and $r_f$ is given by
\begin{equation}
\label{whole_set_rf}
r_f=r_1+\frac {(\pi/2a_v)^{2/3}} {2 k_1^{1/3}}+\frac{a_v(2\pi)^4}{2(r_2-r_1)^2 
r_2^3 k_1^2}.
\end{equation}

Equations (\ref{whole_set_Duvall}-\ref{whole_set_rf}) are the only equations 
needed for the frequency determination. This shows the simplicity of the present 
approach. Three functions of the equilibrium structure are directly involved, 
$c_0 (r)$, $\rho_0 (r)$ and $H_p (r)$. $H_p (r)$ is in fact only important at 
the surface in the determination of $r_2$. $\rho_0 (r)$ contributes via the 
gravitational potential, which plays a crucial role toward the center, being 
more important for lower degree modes.

One can also notice the presence of the coefficient $a_v$ at external as wel as 
internal limit conditions. Determined initially by a `vacuum' condition at the 
surface, it turns to play a key role in the study of the validity of the JWKB 
solution in general.

The above equations are obtained without neglecting the gravity effect. The 
induced results in the following sections will thus present the same quality at 
low as well as at large degree or order, except when deliberately ignoring the 
gravity term.

In the following, the results obtained with the above equations are exposed and 
commented, for the modes $l=0$ to 10, and frequencies in the range 300 - 5600 
$\mu$Hz, firstly for the case without gravity, and secondly in the general case 
taken into account the gravity.

\subsection{Results in the Case Without Gravity}
\label{Results in the Case Without Gravity}

Although neglecting the gravity is a crude approximation, it is very instructive 
because the corresponding physical conditions look like those of acoustic waves 
on Earth, which appear more familiar to us. Furthermore, that will allow direct 
comparisons with the existing semi-analytical approaches where the gravity 
effect is neglected or treated at the second order.

\begin{figure*}[htb]
\centering

\vspace {0.5cm}
\begin {tabular}{cc}
\includegraphics[width=8cm]{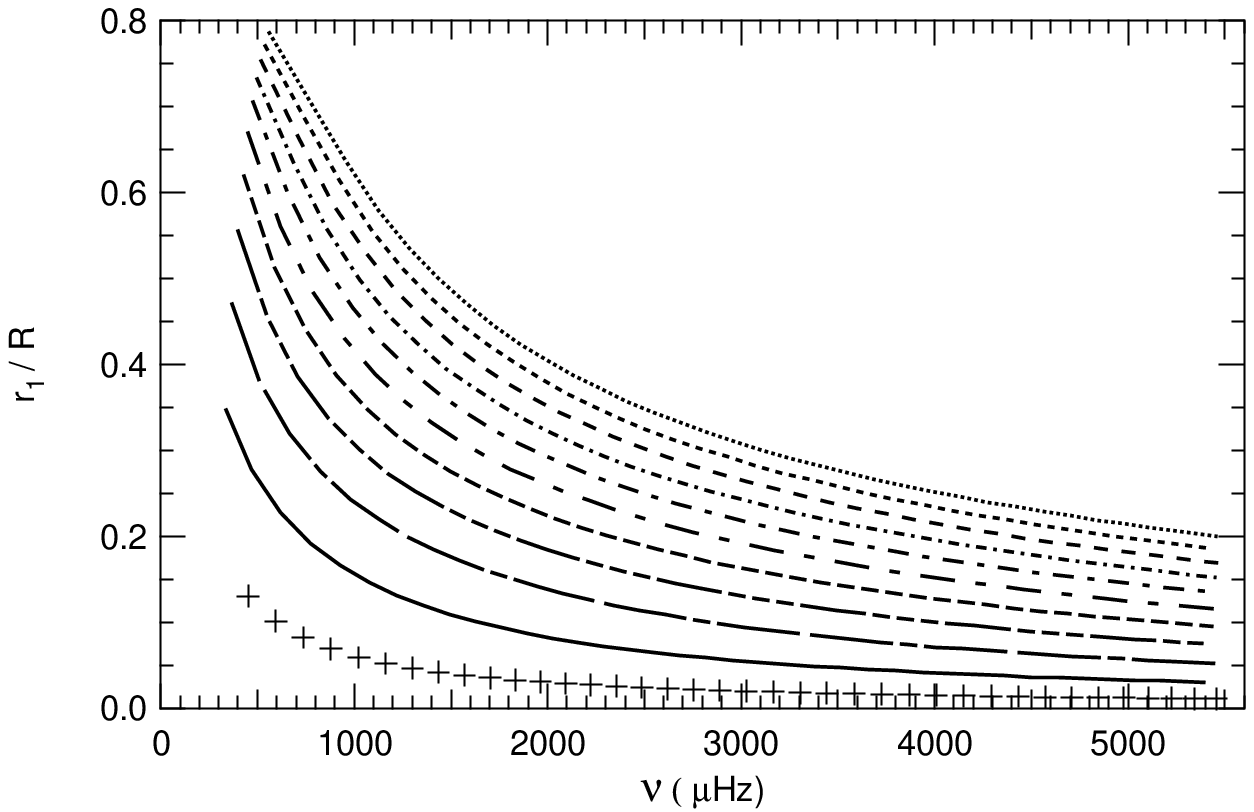} &
\includegraphics[width=8cm]{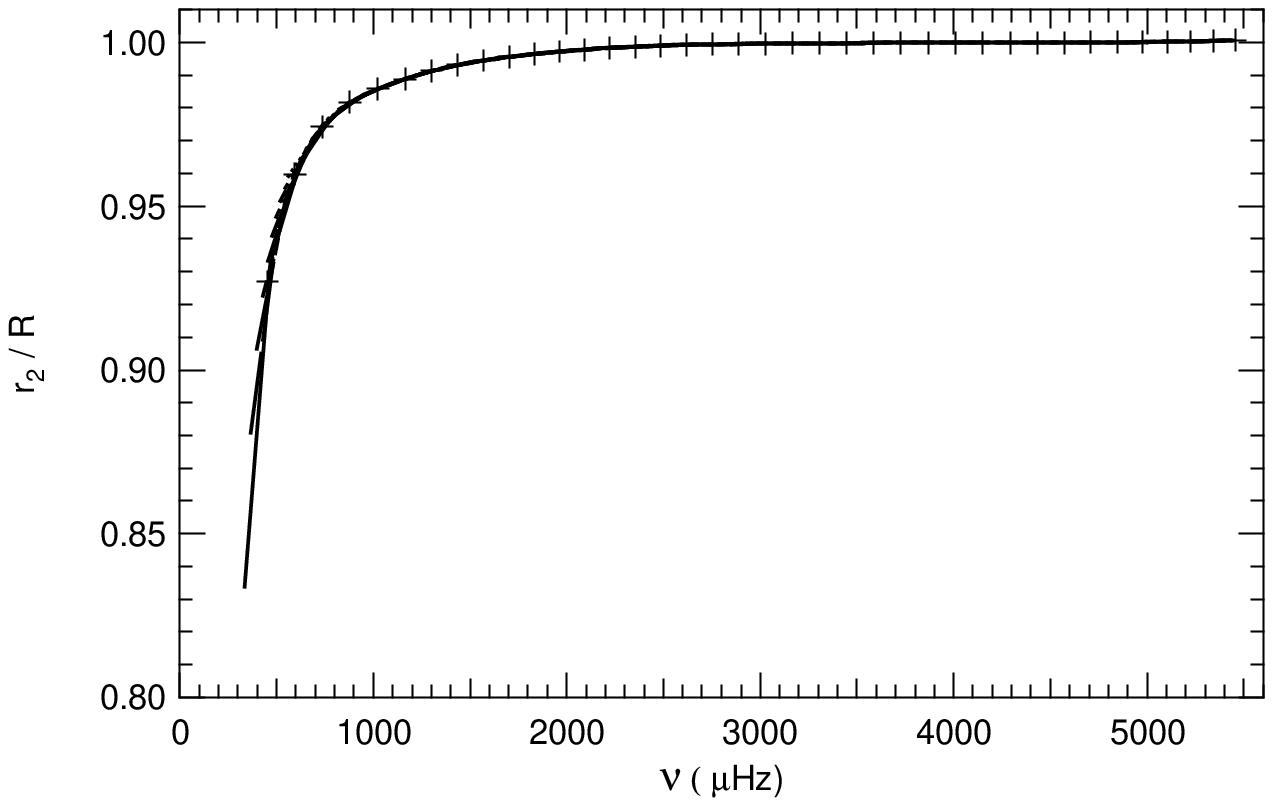}      \\
\includegraphics[width=8cm]{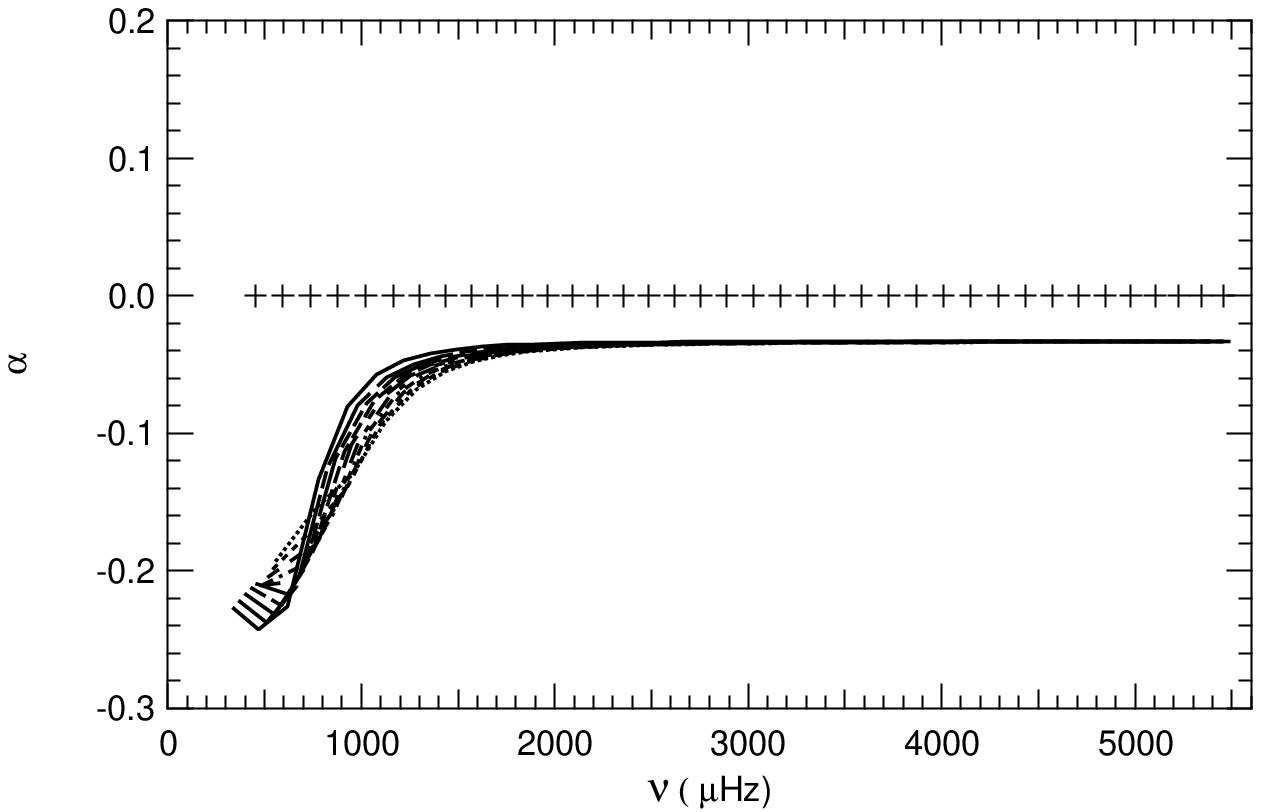} & 
\includegraphics[width=8cm]{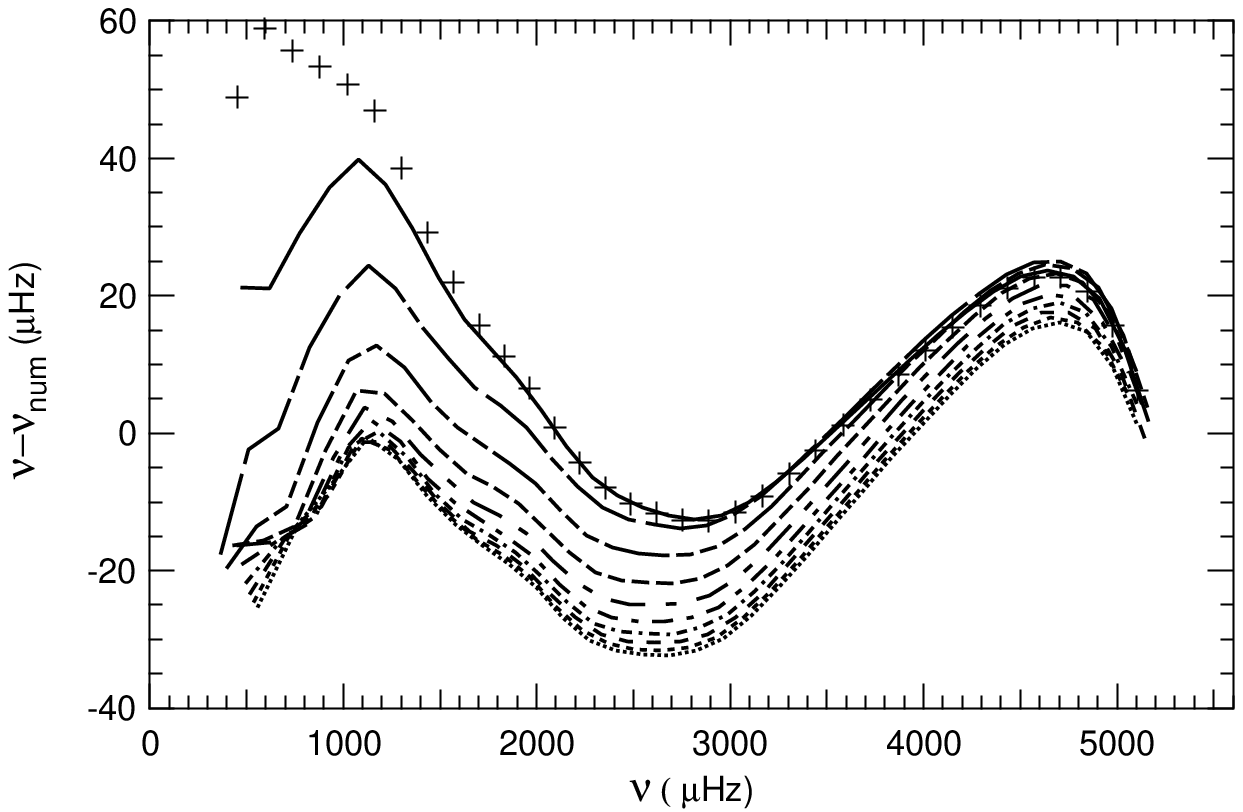} 
\end {tabular}
\caption{Case without gravity. Turning points, phases and frequencies for the 
modes $l=0$ to 10, compared to numerical frequencies where the gravity 
perturbation is taken into account. The symbols are defined in Figure 
\ref{fig_legende}.}
\label{fig_r1_r2_alpha_nu-nunum_asymp}
\end{figure*}

We employ the equations (\ref{whole_set_Duvall}-\ref{whole_set_rf}) with 
$G\rho_0 \rightarrow 0$. The results are displayed in Figure 
\ref{fig_r1_r2_alpha_nu-nunum_asymp} for the turning points $r_1$, $r_2$, the 
phase shift $\alpha$ and the difference $\nu-\nu_{num}$ between the 
eigenfrequencies resulting from the present work and those computed numerically 
(The numerical calculation automatically takes into account perturbation on 
gravity).

%

The phase shift $\alpha$ lies in the range [-0.25, 0]. All the radial modes have 
the same $\alpha=0$, expressing the presence of a `vacuum pressure'. The 
non-radial modes have a common $\alpha$ value only at large frequencies. Indeed, 
for these latter, when $k_1$, the slope of $k_r^2$ ar $r_1$, is enough large, 
the last part of equation (\ref{whole_set_rf}) is ngligible compared to the 
second one, and equation (\ref{whole_set_alpha_lneq0}) gives a constant value 
for $\alpha$
\begin{equation}
\label{alpha_constant}
\alpha=-0.0334
\end{equation}
which is independent of $l$ and $n$.
 
The curves representing the inner turning points $r_1$ can be compared to those 
presented in \citet{Lopes&Turck-Chieze_1994}, obtained from solar observed modes 
\citep{Libbrecht_etal_1990}, with a formalism using $L=(l+0.5)^2$. While given 
by diffferent formulas, the $r_1$ of radial and non-radial modes seem to be very 
similar.

The outer turning points $r_2$ feature a remarkable behaviour: for frequencies 
beyond 800 $\mu$Hz, $r_2$ depends very few on the mode order $l$ and much more 
on frequency. Indeed, equations (\ref{whole_set_kr}) and (\ref{whole_set_r2}) 
show that for low $l$, $k_r$ and $r_2$ are independent of $l$. That means that a 
modification of the Sun surface will modify in the same way all the modes of 
different degree. This fact will be exploited hereafter. This set of very 
grouped curves represents the external profile of the Sun acoustic cavity. One 
can also remark that $r_2$ increases very quickly with frequency, right beyond 
2500 $\mu$Hz, its curve reaches already the flatten part, very close to 
$r_2/R=1$. The results from 1000 $\mu$Hz can be compared to the curve obtained 
by \citet{Deubner&Gough_1984} (and detailed in \citet{JCD_cours_1998}), 
following a general asymptotic expression obtained after several variable 
changes. One has otherwise keep in mind that:
\begin{enumerate}
  \item Confusing $r_2$ with $R$ leads to frequency displacements of tens of 
$\mu$Hz.
  \item For larger mode degrees, $r_2$ depends of course stronger on $l$ 
following equation (\ref{whole_set_r2}).
\end{enumerate}

Concerning the comparison to numerical results for the eigenfrequencies, a first 
observation of these raw data leads to a rather disappointing report: the 
differences reach -32/+60 $\mu$Hz. But a more detailed analysis shows that the 
differences to be taken into account are much smaller. Actually, these 
discrepancies consist mainly in a global undulation independent of $l$, 
signature of a difference, solely at the surface, between the two calculations. 
As justified in the next sections, a simple correction of surface conditions to 
be in accordance with the numerical code will cancel this undulation. Therefore 
only the thickness of the set of these curves is to be taken into account. 
Straight away then, the classical asymptotic character clearly appears, the 
differences being progressively bigger when the frequencies go lower, according 
to the importance of $G\rho_0$ in the expression of $k_r$ (eq. 
\ref{whole_set_kr}). By discarding the results at very low frequencies where the 
coupling with gravity modes can no more be neglected, in all the rest between 
800  and 5100 $\mu$Hz, the differences are reduced to no more than $\pm34\mu$Hz 
to $\pm4\mu$Hz. Although having different $r_1$ and $\alpha$, the radial modes 
present differences with numerical results which are in continuity with the 
non-radial ones.

\subsection{Results in the General Gase (With Gravity)}
\label {Results in the General Case}

\begin{figure*}[htb]
\centering
\vspace {0.5cm}
\begin {tabular}{cc}
\includegraphics[width=8cm]{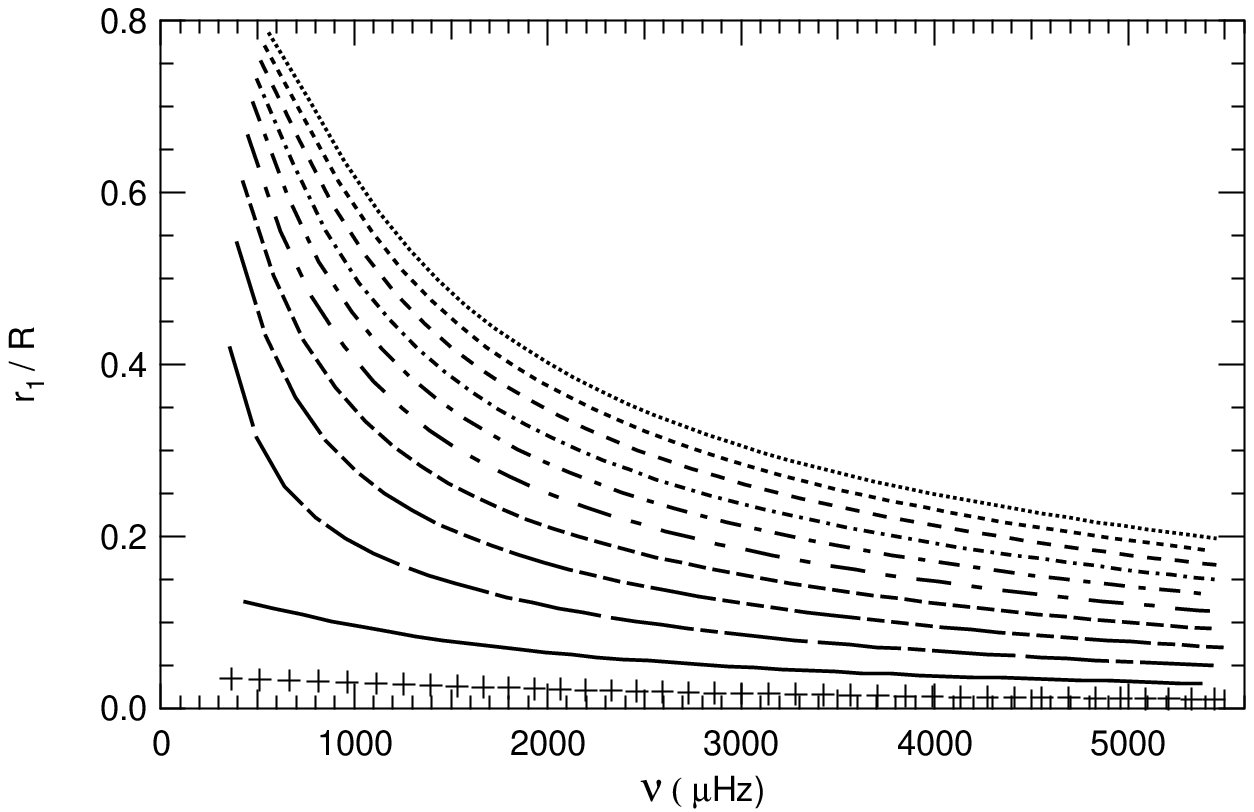} &
\includegraphics[width=8cm]{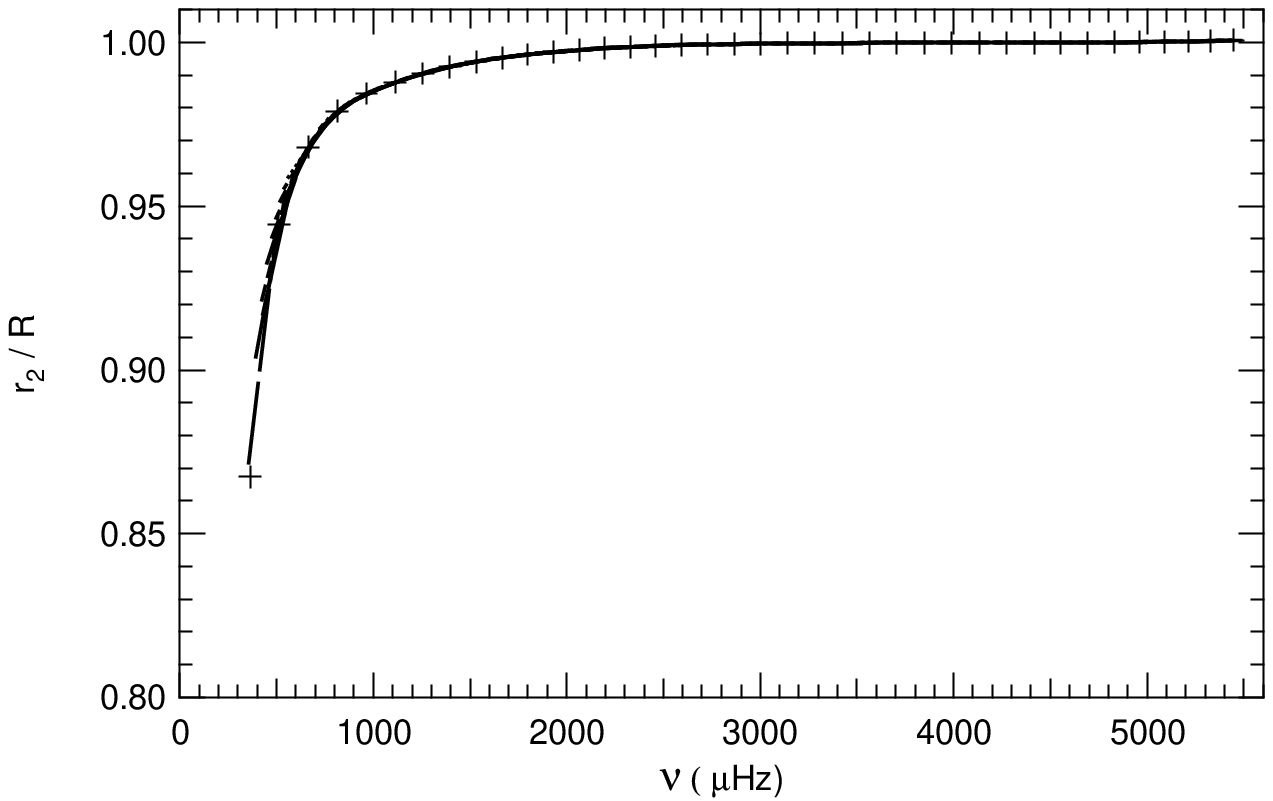}      \\
\includegraphics[width=8cm]{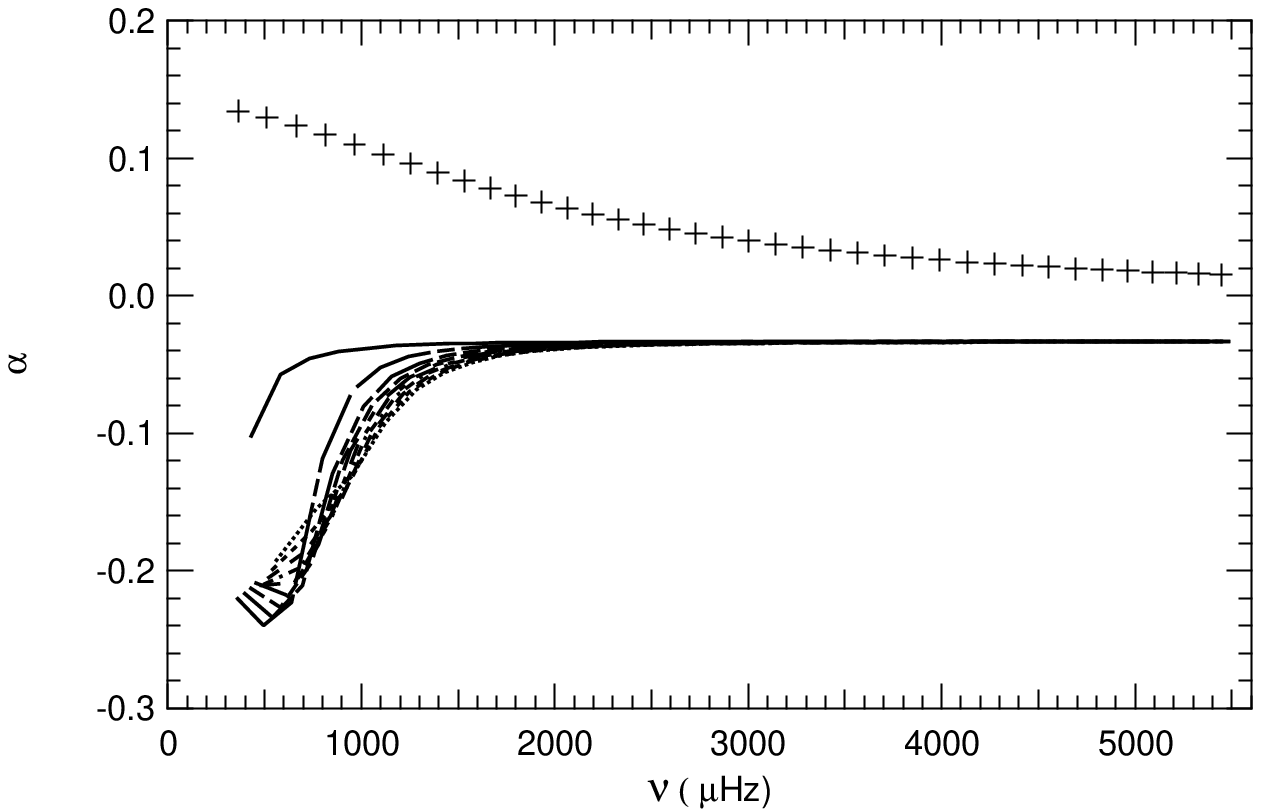} & 
\includegraphics[width=8cm]{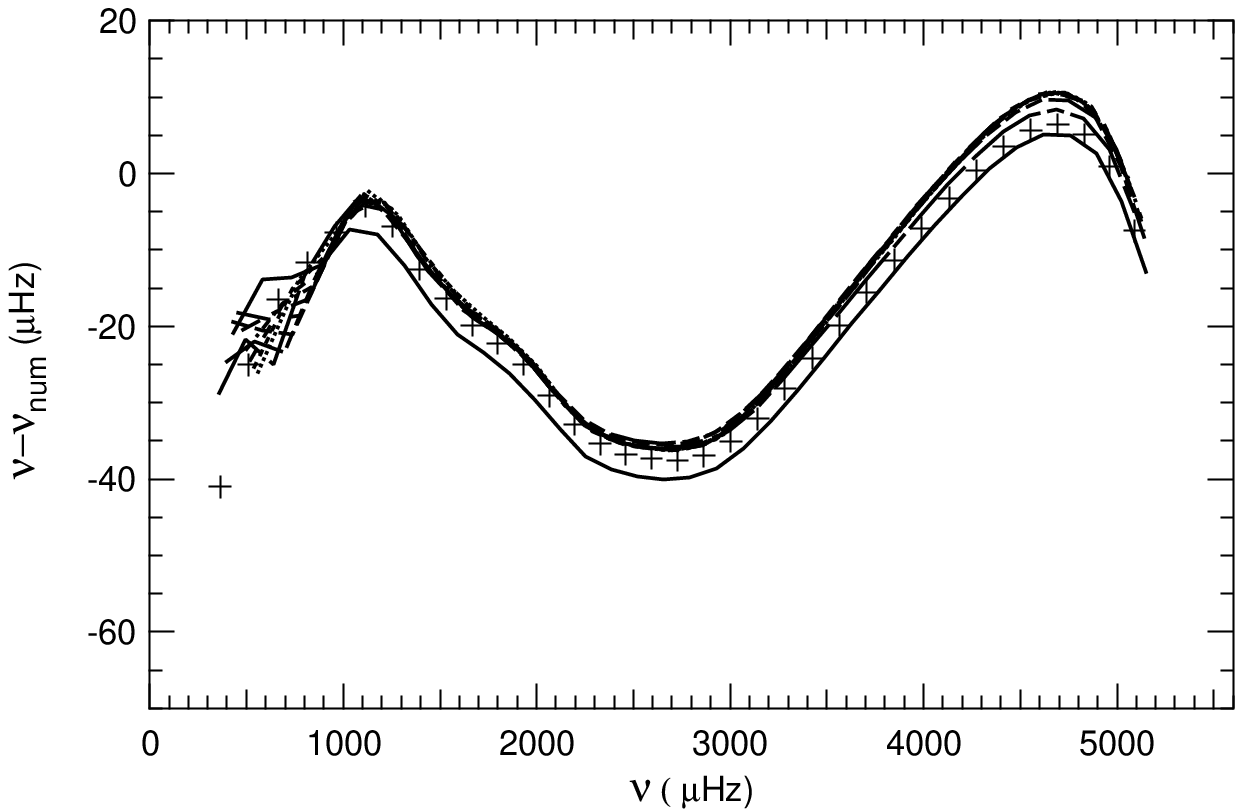} 
\end {tabular}
\caption{General case in the presence of gravity. Turning points, phases and 
frequencies for the modes $l=0$ to 10. The symbols are defined in Figure 
\ref{fig_legende}.}
\label{fig_r1_r2_alpha_nu-nunum}
\end{figure*}

We use the set of equations (\ref{whole_set_Duvall}-\ref{whole_set_rf}) as they 
are. The results are displayed in Figure \ref{fig_r1_r2_alpha_nu-nunum}.

The external turning points $r_2$ are the same than without gravity because the 
gravity does not play any role at the surface. Every related comment in the 
previous section remains totally valid here.

As foreseen, the internal turning points $r_1$ are especially smaller for 
low-degree, low-frequency modes, the gravity acts like if it pulls these modes 
toward the centre. So the realted $\alpha$, which depends on $r_1$, is smaller 
in absolute value. These features are particularly noticeable  for the $l=0$ and 
$l=1$ modes. The common value of $\alpha$ at large frequency, $\alpha$=-0.0334, 
remains the same. The $\alpha$ of the radial modes are now completely different, 
and are the only ones that are positive.

These different $r_1$ and $\alpha$, together with different $k_r$ in the inner 
parts will induce different eigenfrequencies when gravity is taken into account. 
One can moreover predict that all the frequencies will be lowered. These 
differences are larger for lower degree or higher frequency modes which travel 
deeper. To appreciate that numerically, some examples of frequencies without or 
with gravity, are given in Table \ref{table_freq} for various degrees and 
orders. The same type of comparison for $r_1$ was given in Table \ref{table_r1}.
\begin{table}[htb]
\caption{Frequencies in $\mu$Hz without / with gravity, for various degrees and 
orders.}
\label{table_freq}
$ \begin{array} {lccc}
\hline
        &      n=2       &           n=10           &       n=30      \\
\hline
l=0     & 453.2 / 363.7  &\ \  1570.1 / 1531.7\ \   & 4291.8 / 4273.6 \\
l=1     & 469.6 / 427.4  &\ \  1629.0 / 1591.3\ \   & 4360.7 / 4340.8 \\
l=2     & 512.2 / 492.7  &     1680.9 / 1655.6      & 4427.7 / 4410.0 \\
l=5     & 622.1 / 618.8  &     1810.9 / 1801.8      & 4601.3 / 4590.5 \\
l=10    & 767.3 / 766.4  &     1998.5 / 1995.7      & 4840.9 / 4835.4 \\
\hline
\end{array} $
\end{table}
%


%

The curves of Figure \ref{fig_r1_r2_alpha_nu-nunum} featuring the difference 
with numerical calculation show the same global undulation independent of $l$, 
at exactly the same locations in radius than with the case without gravity. For 
the same reason, only the thickness of this set of curves is to be considered. 
One can then conclude that between 800 and 5100 $\mu$Hz, the discrepancy is 
drastically reduced to within $\pm$ 3 $\mu$Hz. And the most remarquable point is 
that the former asymptotic behaviour when the gravity was neglected, has 
disappeared. Even radial and non-radial modes, which are governed by different 
physics of internal boundary condition, present the same profiles. This can lead 
to attribute a higher degree of confidence to the present calculations in 
general, and to the coefficient $a_v$ in particular.

\subsection{Discussion on the Outer Boundary Conditions}
\label{Discussion on the Outer Boundary Conditions}

The gravity being taken into account, it is now worth examining the global 
undulation independent of $l$ (Fig. \ref{fig_r1_r2_alpha_nu-nunum}) 
characterising the frequency differences with numerical results. As the solar 
model used is the same, that points out a difference in the external boundary 
conditions used in the two frequency calculations: the present semi-analytical 
calculation applies the condition of local inhomogeneity, while the numerical 
code is run with the isothermal atmosphere approximation.

Although the isothermal atmosphere approximation leads to frequencies that are 
closer to observations, several non-satisfying points remain:
\begin{itemize}
\item It is not very realistics, at least when compared to a semi-empirical 
atmosphere of the type of \citet{VAL_1981}, which presents no place with a flat 
temperature profile. No atmosphere coming from a stellar evolution code presents 
any more such a behaviour at the surface.
\item It is not really physics, because an infinite isothermal atmosphere needs 
to be invoked, whose integrated mass is infinite.
\item It requires to adjust a particular atmosphere to the input stellar 
model, thus actually changes the initial surface conditions of the latter. When 
discrepancies with observations are detected, it is not possible to distinguish 
the part coming from the initial surface conditions, and the part coming from 
its change by the adjustment of an isothermal atmosphere.
\end{itemize}

\begin{figure}[htb]
\centering
\plotone {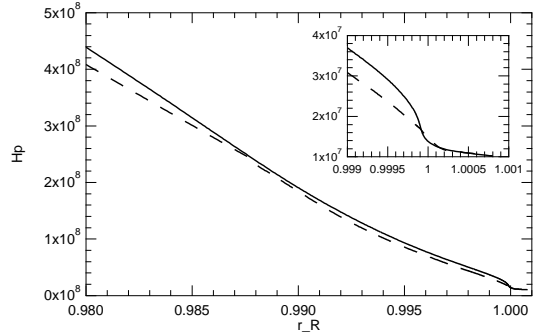}
\caption{Profiles of the pressure scale height at the surface and zoom at the 
external limit. Continuous line: initial $H_p$ coming from the stellar evolution 
code. Dashed-line: $H_p$ expected to come from the double adjustment of the 
isothermal atmosphere approximation.
        }
\label{fig_Hp_num}
\end{figure}

A search for boundary conditions using directly the input stellar structure 
without any modification, may be desirable. Then, any difference between two 
sets of frequencies could be attributed to differences in the stellar structures 
considered. The condition proposed here (eq. \ref{whole_set_r2}) is only a 
suggestion in this direction, which  potentially allows to recover surface 
properties. To prove that, we can try here to find out the amosphere properties 
used by numerical calculation. As the wave solution is adjusted to that of the 
analytical solution in an infinite isothermal atmosphere adjusted to the initial 
atmosphere, this double adjustment amounts to modify very deeply the profiles of 
$\Gamma_1$ and $H_p$ while keeping unchanged their product. So the wave 
propagation is not affected, only its outer reflexion is concerned. Therefore we 
have to find out the $H_p$ change that cancel the frequency discrepancies. In 
our case, this is simple to do because there is a one-to-one correspondence 
between $\nu$ and $r_2$ (upper right panel of Fig. 
\ref{fig_r1_r2_alpha_nu-nunum}) and between $r_2$ and $H_p$ (Fig. 
\ref{fig_ex_r2}), thus between $\nu$ and $H_p$. The result is shown in Fig. 
\ref{fig_Hp_num}. In \citet{Nghiem_2003b}, it was demonstrated that this profile 
is actually the $H_p$ profile when the procedure of isothermal atmosphere 
adjustment is performed. This profile does not have any consistency with the 
stellar structure because it results from a fit with a structure having very 
different properties.

\begin{figure}[htb]
\centering
\plotone {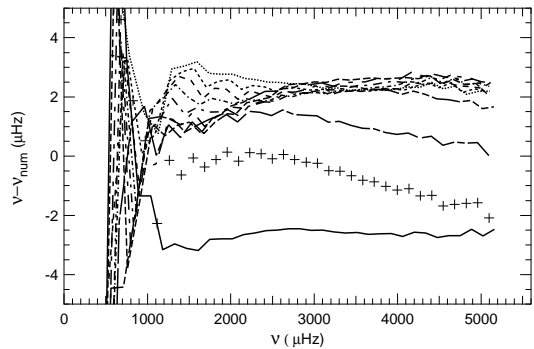}
\caption{Resulting frequency discrepancy with numerical results for the modes 
$l=0$ to 10, in the presence of the $H_p$ profile expected to come from the 
isothermal atmosphere approximation (Fig. 
\ref{fig_Hp_num}). The symbols are defined in Fig. \ref{fig_legende}.
        }
\label{fig_nu-nunum_res}
\end{figure}

With this deduced $H_p$ profile, the final frequency discrepancy is shown in 
Fig. \ref{fig_nu-nunum_res}. It remains large only for low frequencies $<$ 800 
$\mu$Hz, where coupling with gravity modes is no more negligible. For 
frequencies between 800 and 5100 $\mu$Hz, the discrepancy is on the contrary 
reduced as expected to less than $\pm 3\mu$Hz.



%

\section{Comparison With Observations}
\label{Comparison With Observations}

\begin{figure}[htb]
\centering
\plotone {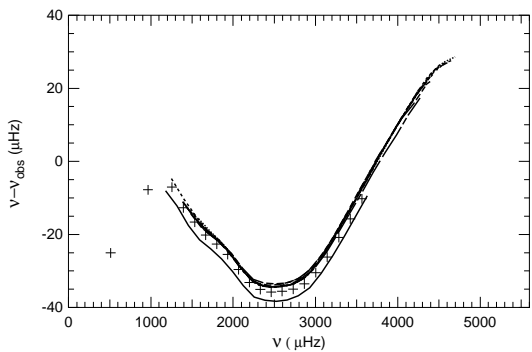}
\caption{Frequency differences with observational data for the modes $l=0$ to 
10. The symbols are defined in Fig. \ref{fig_legende}.
        }
\label{fig_nu_nuobs}
\end{figure}

\begin{figure}[htb]
\centering
\plotone {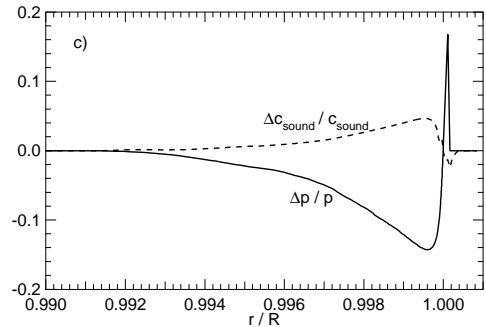}
\caption{Profiles at the surface of the pressure and sound speed changes, 
expected to come from a magnetic field, which cancel discrepancies with 
observational frequencies.
        }
\label{fig_dc_dp}
\end{figure}

\begin{figure}[htb]

\centering
\plotone {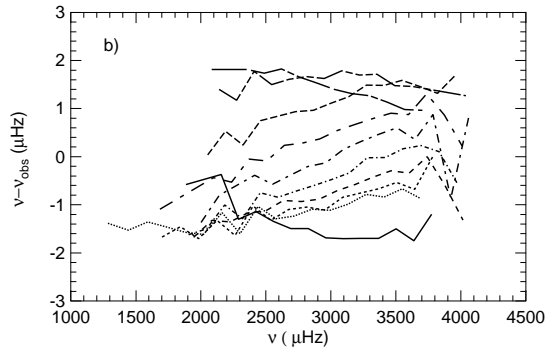}
\caption{Resulting frequency discrepancies with observational data for the modes 
$l=1$ to 10, after the changes of pressure and sound speed in Fig. 
\ref{fig_dc_dp}. The symbols are defined in Fig. \ref{fig_legende}.
        }
\label{fig_nu_nuobs_res}
\end{figure}

Fig \ref{fig_nu_nuobs} gives the differences between frequencies obtained by the 
present study on the solar structure coming from the evolution code, and those 
obtained by solar observations coming from the SoHO spacecraft. There is a 
global modulation independent of $l$ like with numerical results but different 
in values, going down to -40 $\mu$Hz, and up to +30 $\mu$Hz. This comes 
indisputably from problems at the surface, and can be attributed to the fact 
that the solar model used does not yet take into account some important physical 
phenomenon. In \citet{Nghiem_etal_2004b} it was assumed that it concerns 
exclusively a magnetic field, which modifies the structure  by the magnetic 
pressure and the Alfven speed. Then we aim to search for perturbed pressure and 
sound speed profiles that cancel the frequency discrepancies. The results are 
shown in Figure \ref{fig_dc_dp}. That finally allows to obtain the frequency 
differences of only $\pm 2 \mu$Hz in the whole studied frequency range (Fig. 
\ref{fig_nu_nuobs_res}). In the same way, a magnetic variation has also been 
calculated to account for the frequency variation related to the solar cycle. 
But this study has to be confirmed by a more detailed one, especially by taking 
into account the density change with the introduction of the magnetic field.

These results does not exclude that other usual phenomena (turbulent pressure, 
non-adiabaticity, etc.) can reduce the theory-observation discrepancy in 
frequencies. Thay have only shown the capabilities of the proposed method to 
recover identified surface properties, because no a-priori hypothesis was made 
on the structure. If numerical codes can adopt the same type of boundary 
conditions than here, absolute frequency values could be calculated at a 
precision well better than $\pm 1\mu$Hz in the whole frequency range. That will 
be very helpful for asteroseismology in the identification of the modes, when 
only very few low-order frequencies will be available.


\section{Conclusions-Discussions}

The local wave approximation has been fully and coherently applied to stellar 
acoustic modes. Conceptually, it is in total accordance with the JWKB approach. 
Pratically, it consists in clearly determining the cavity limits and the 
reflection conditions. The standing wave nature of the modes are thus firmly 
claimed. In this simple context, a particularly reduced number of equations is 
necessary.

The wave propagation in the innner part is nevertheless correctly represented, 
as the gravity effect is directly taken into account. No development into higher 
orders, no artificial change of the horizontal eigenvalue $l(l+1)$ are needed. 
It turns out that the $l=0$ modes do not touch the centre. Not surprisingly, the 
gravity plays an important role for low-$l$ modes, being able to shift 
frequencies by tens of $\mu$Hz. The main result is that the present analytical 
calculation is equally valid for low than high frequencies and degrees, only the 
coupling with gravity modes is not considered.

At the surface, a new boundary condition is proposed, directly based on the 
propagation condition of an acoustic wave, which corresponds to the JWKB 
validity criteria. Although that returning point should perhaps be refined, some 
first conclusions can already be drawn. A coefficient like $a_v$ seems to be 
sufficient to characterize the validity of a JWKB solution. But the main result 
is that an alternating boundary condition leads to frequency shifts of the same 
range than present theory-observation discrepancies. To explain the latter, a 
search for a more appropriate boundary condition cannot be excluded. The method 
proposed here has the advantage to use the atmosphere model as it is, without 
any modification. Comparisons with observations should allow to directly recover 
the real atmosphere structure. Examples are given to find out the atmosphere 
used by numerical codes in the context of isothermal atmosphere approximation, 
and to suggest a way to deduce physical properties of the solar atmosphere when 
observations are considered.

Despite the simplicity of the proposed semi-analytical formalism, it seems that 
main physical phenomenons involved are taken into account. That is why for the 
Sun, a precision of $\pm 3 \mu$Hz can be reached in the range 800-5600 $\mu$Hz, 
and for $l=0-10$. Some calculations for other stars show a comparable behaviour. 
If now a similar type of external boundary conditions is used in numerical 
calculations, a potential precision well better than 1$\mu$Hz can be expected 
for the whole frequency range. That could represent a step forward a global 
agreement between the three actors involved in helio-, astero-seismology: 
semi-analytical calculation bringing the physical understanding by predicting 
the good trend for the results, numerical calculation bringing the high 
precision by avoiding most of the approximations, and observation allowing to 
infer real stellar structures when searching to cancel theory-observation 
discrepancies.

\begin{acknowledgements}
\emph{Acknowledgements.} I am indebted to S. Turck-Chi\`eze  for her precious remarks and her support, to 
R. A. Garc\'\i a,  S. Couvidat, L. Piau, J. Ballot, and S. Brun for many 
fruitful discussions. They all have helped me to clarify the physical points 
presented here.
\end{acknowledgements}



\appendix

\section{The JWKB Approach}
\label{The JWKB Approach}

Only a brief recall of the JWKB method is proposed here, detailed historical and 
theoretical considerations are out of this paper frame. The reader can see e.g. 
\citet*{Elmore&Heald_1969}. Let $y$ be the unknown function of the variable $r$ 
ruled by the propagation equation
\begin{equation} \label{wave_eq_kconst}
\frac{d^2y}{dr^2}+ k_r^2 y=0
\end{equation}
with $k_r$ a constant. The solution is of course the wave function:
\begin{equation} \label{wave_fct_kconst}
y=A\exp (\pm ik_r r)
\end{equation}
with A an arbitrary constant. Consider now the case where $k_r$ is itself a 
function of $r$, the equation
\begin{equation} \label{wave_eq_kvar}
\frac{d^2y}{dr^2}+ k_r^2(r)y=0
\end{equation}
admits no known solution in the general case, and certainly no wave solution 
like equation (\ref{wave_fct_kconst}). A complex solution of the type
\begin{equation} \label{wave_fct_kvar_tried}
y=A(r)\exp [iS(r)]
\end{equation}
can be tried, with A, S functions of $r$. By separating real and imaginary 
parts, one obtains:
\begin{eqnarray} 
\label{solution_kvar_A}
 A&=&CS'^{-1/2} \\
\label{solution_kvar_S'}
 S'^2 &=& k_r^2-\frac{1}{2}\frac{S'''}{S'}+\frac{3}{4}\frac{S''^2}{S'^2}
\end{eqnarray}
where $S',S'',S'''$ stands for the successsive derivatives of $S$. When $S'''$ 
and $S''$ are negligible compared to $S'$, 
\begin{equation} \label{JWKB_cond}
\frac{S''^2}{S'^2}<<k_r^2 \ \ \ \ and \ \ \ \frac{S'''}{S'}<<k_r^2
\end{equation}
and only in that case, equation (\ref{solution_kvar_S'}) can be simplified to
\begin{equation} \label{solution_S'_approx}
S'=\pm k_r
\end{equation}
and one has a wave-like solution:
\begin{equation} \label{wave_fct_JWKB}
y=\frac{C}{\sqrt{k_r}}\exp\left(\pm i\int_{}^{}\!\! k_r dr \right)
\end{equation}
with C an arbitrary constant.

\section{Eigenfunction, Radial Density of Kinetic Energy, and $k_r$}
\label{Eigenfunction}

There is sometimes a  confusion between the eigenfunction and the radial density 
of kinetic energy, or related quantities... Let us first of all precise that 
here the eigenfunctions are of the type $y(r)$ given by Eq. 
(\ref{standing_wave}), where $y$ stands either for $\rho'$, $p'$, $\phi'$, or 
$\xi_r$.

\begin{figure}[htb]
\begin {center}
\begin {tabular}{c}
\includegraphics[width=8cm]{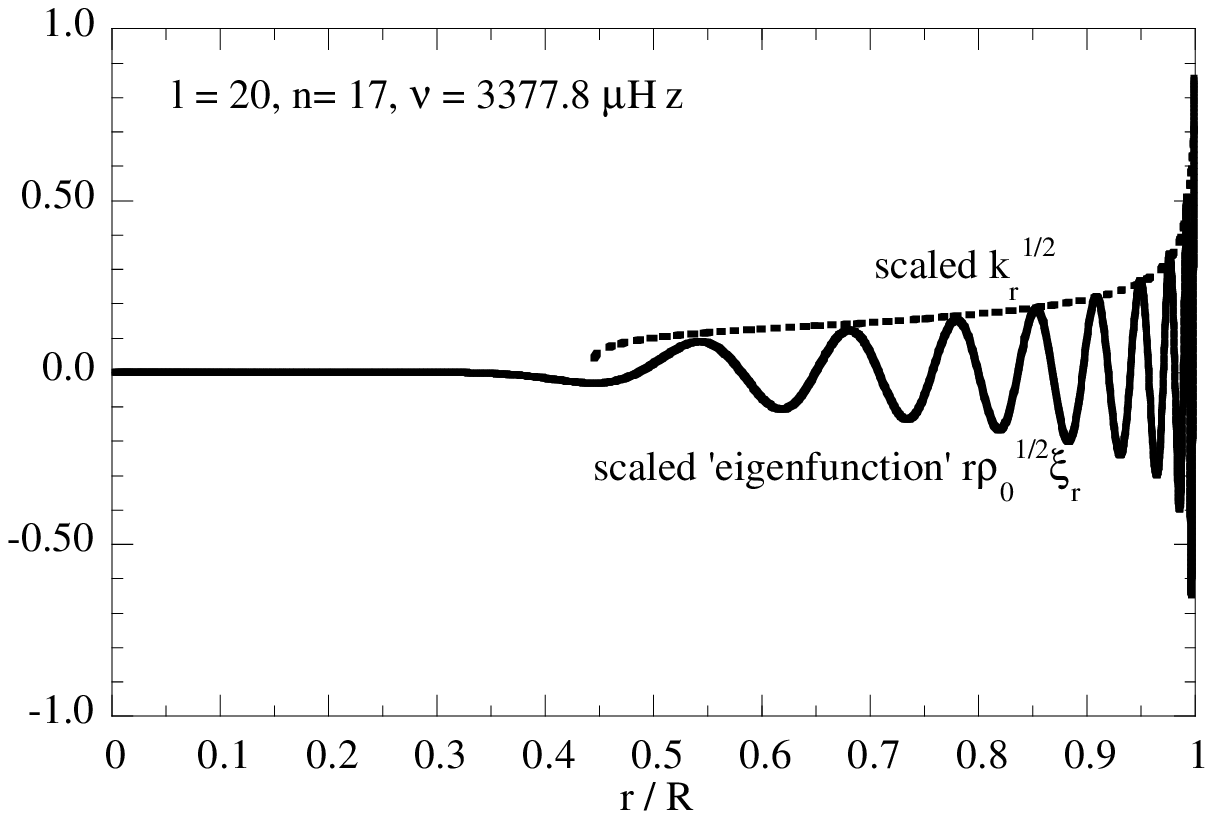} \\                                                                                                                                                                                                                                                                        
\includegraphics[width=8cm]{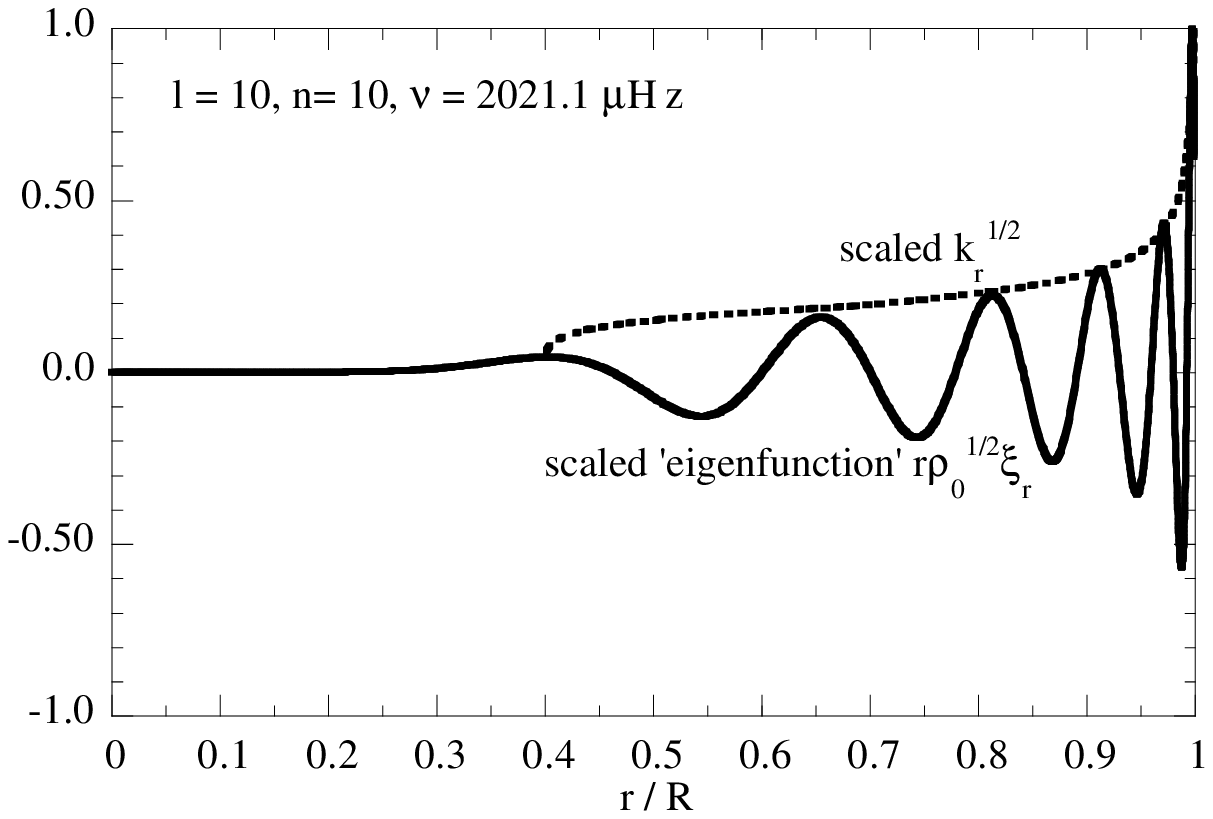} \\
\includegraphics[width=8cm]{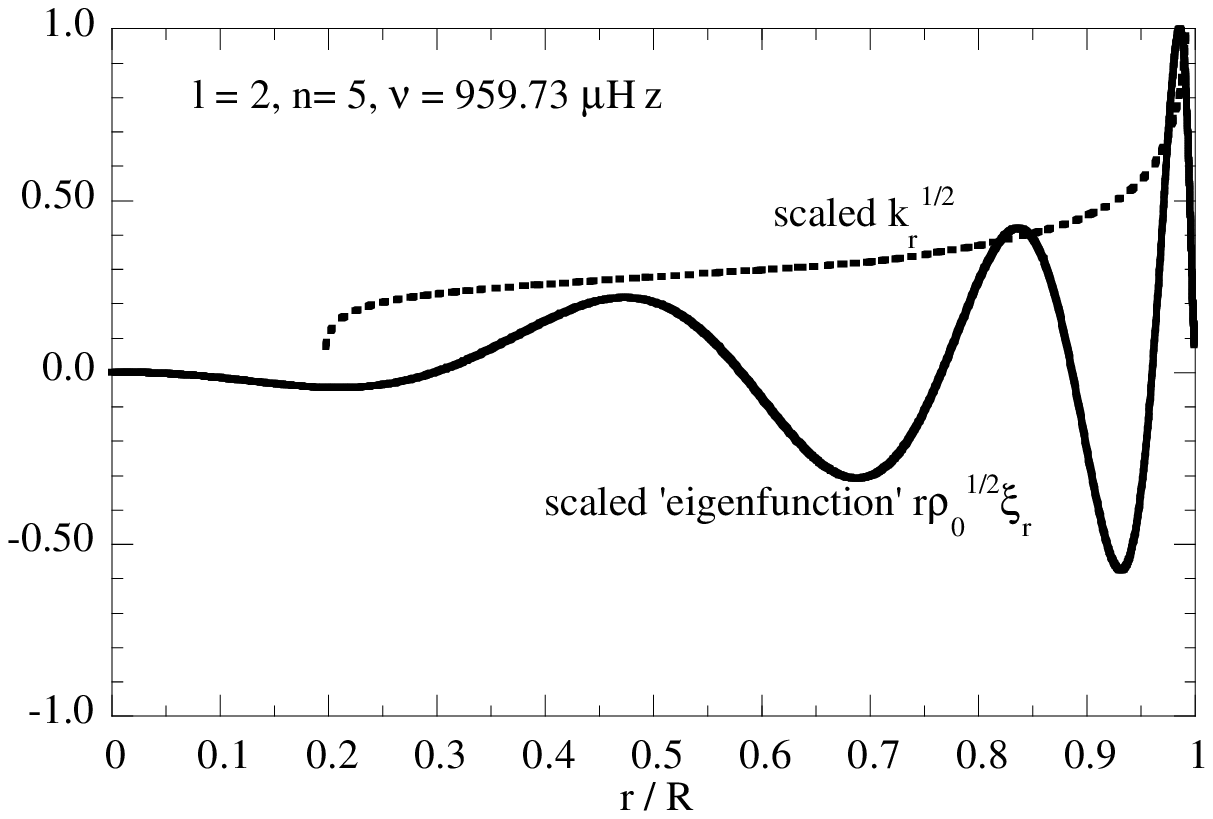}  
\end {tabular}
\caption{Scaled $k_r^{1/2}$ (dashed-line) superimposed on scaled 
`eigenfunctions' (continuous line) coming from the numerical code, for three 
different cases of $l$, $n$.}
\end {center}
\label{fig_pseudo_eigenfunction}
\end{figure}

The amplitudes $A/r$ remain to be determined. It can be done in the context of 
the JWKB approximation (eq. \ref{wave_fct_JWKB}), but that is not enough 
detailed because some slowly varying terms can still be contained in the 
`constant' amplitude that is infact slowly variable itself. For $\xi_r$ for 
example, physical considerations about kinetic energy allow to have a more 
detailed determination of the amplitude. For a locally plane wave, the kinetic 
energy integrated in each wavelength (one spatial oscillation) is the same. So 
the kinetic energy per unit length is proportional to the number of wavelengths 
per unit length, i. e. the wavenumber:
\begin{equation} 
\frac{\partial E_k}{\partial s} \propto k
\end{equation}
where s is the curviligne coordinate following the wave trajectory. Now for the 
motion of a locally spherical wave projected on the radial direction, the 
quantity that is conserved and proportional to $k_r$ is the radial kinetic 
energy per unit of radius, or if one prefers, its flux over any sphere of radius 
$r$:
\begin{equation}
\frac{\partial \epsilon_k}{\partial r} = \frac{1}{2}\rho_0\left(\frac{\partial 
\xi_r}{\partial t}\right)^2 r^2 = \frac{\omega^2}{2}\rho_0\xi_r^2 r^2 \propto 
k_r
\end{equation}
So the amplitude of $\xi_r$ is proportional to
\begin{equation}
\xi_r \propto \frac{\sqrt{k_r}}{r\omega\sqrt{\rho_0}}.
\end{equation}
That is the reason why sometimes $r\sqrt{\rho_0}\,\xi_r$ is called the 
`eigenfunction', which is in fact proportional to the square root of 
$\partial\epsilon_k/\partial r$. Figures \ref{fig_pseudo_eigenfunction} show 
that this quantity is well proportional to $\sqrt{k_r}$. Some other times, that 
is rather  $r\sqrt{\rho_0 c_0}\,\xi_r$ which is called the `eigenfunction', and 
this function has consequently a constant amplitude along $r$.

%



\end{document}